\documentclass[aps,prb,twocolumn,groupedaddress]{revtex4-1}

\usepackage{graphicx}
\usepackage{amsmath}
\usepackage{amsfonts}
\usepackage{braket}
\usepackage{color}
\definecolor{Blue}{rgb}{0.3,0.3,0.9}
\definecolor{Red}{rgb}{0.9,0.3,0.3}

\begin{document}


\title{Self-assembling tensor networks and holography in disordered spin chains}


\author{Andrew M.\ Goldsborough}
\email[]{a.goldsborough@warwick.ac.uk}
\homepage[]{www.warwick.ac.uk/andrewgoldsborough}
\affiliation{Department of Physics and Centre for Scientific Computing, The University of Warwick, Coventry, CV4 7AL, United Kingdom}

\author{Rudolf A.\ R\"{o}mer}
\email[]{r.roemer@warwick.ac.uk}
\homepage[]{www.warwick.ac.uk/rudoroemer}
\affiliation{Department of Physics and Centre for Scientific Computing, The University of Warwick, Coventry, CV4 7AL, United Kingdom}


\date{$Revision: 1.59 $, compiled \today}

\begin{abstract}
  We show that the numerical strong disorder renormalization group
  algorithm (SDRG) of Hikihara et.\ al.\ [{Phys. Rev. B} {\bf 60},  12116
  (1999)] for the one-dimensional disordered Heisenberg
  model naturally describes a tree tensor network (TTN)
  with an irregular structure defined by the strength of the
  couplings. Employing the holographic interpretation of the TTN in Hilbert space,
  we compute expectation values, correlation functions and the entanglement
  entropy using the geometrical properties of the TTN. We find
  that the disorder averaged spin-spin correlation scales with the average path
  length through the tensor network while the entanglement entropy 
  scales with the minimal surface connecting two regions. Furthermore, the entanglement 
  entropy increases with both disorder and system size, resulting in an 
  area-law violation. 
  Our results demonstrate the usefulness of a self-assembling TTN approach to disordered systems and quantitatively validate the connection between holography and quantum many-body systems. 
 \end{abstract}


\maketitle


\section{Introduction}

There is currently a lot of excitement around the so-called AdS/CFT correspondence and possible
applications in condensed matter physics.\cite{Mcg10} The AdS/CFT correspondence 
is most well known in high energy physics where it was noted \cite{Mal99} that
there exists a duality between certain theories of gravity on $D+1$ dimensional
Anti de Sitter (AdS) spacetime and conformal quantum field theories (CFT) living on its $D$ dimensional
boundary. In condensed matter systems, the AdS/CFT correspondence can provide a geometric interpretation of 
renormalization group (RG) techniques since the additional \emph{holographic} dimension can be interpreted as a 
scale factor in the RG coarse graining.\cite{Mcg10} 
It has been argued recently\cite{Swi12,Mol13} that certain RG approaches to the Hilbert space of critical many-body interacting system in $D$ dimensions, such as the multi-scale entanglement renormalisation ansatz (MERA) for tensor networks, share many of their geometric properties with $D+1$ dimensional AdS. This connection is based on ideas\cite{RyuT06} that suggest that the entanglement entropy of a region on the boundary is related to the minimal surface in the holographic bulk that separates the region from the rest
of the surface. These ideas
were further developed by Evenbly and Vidal\cite{EveV11} to discuss the underlying geometric structure of
entanglement and correlation functions in such tensor networks in general. 

Tensor network methods provide
elegant and powerful tools for the simulation of quantum many-body
systems. Their original manifestation, the density-matrix renormalization group (DMRG)\cite{Whi92} is now understood to be
based on a variational update of a matrix product state (vMPS),\cite{OstR95,Sch11} and has 
found applications in a wide range of fields such as quantum chemistry\cite{ShaC12}
and quantum information\cite{PerVWC07} as well as condensed matter physics.\cite{CraFMS08}
More recent developments have extended the methods to, e.g., critical
systems,\cite{Vid07} two-dimensional lattices\cite{VerC04,EveV10,XieCQZ12} and 
topologically ordered states.\cite{DusKOS11}

\begin{figure}[bt]
(a)\includegraphics[width=0.95\columnwidth]{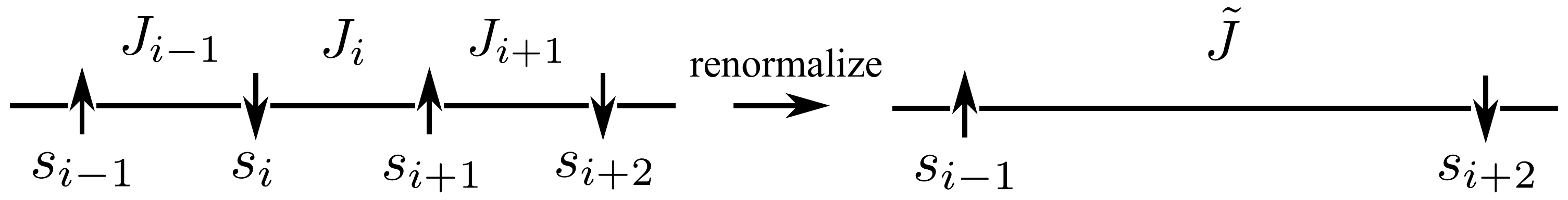}
(b)\includegraphics[width=0.95\columnwidth]{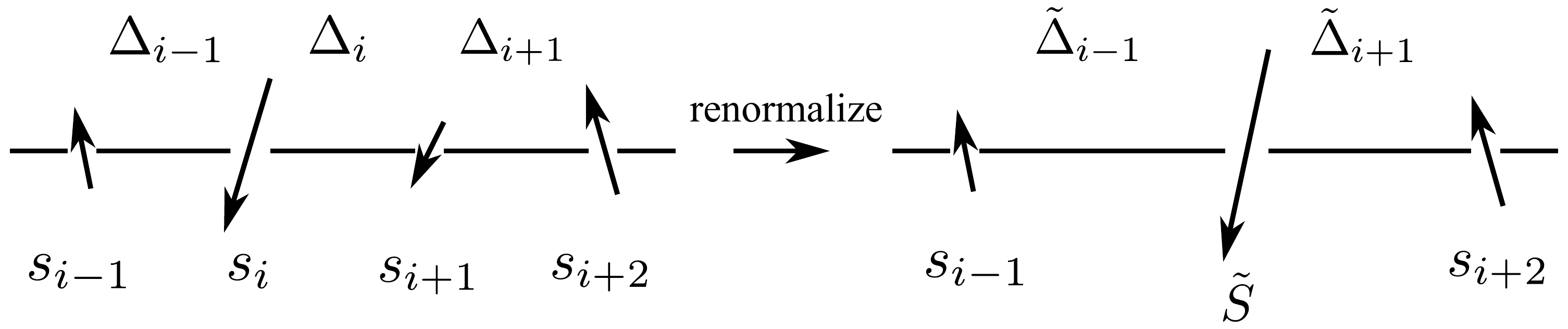}
(c)\includegraphics[scale=0.3]{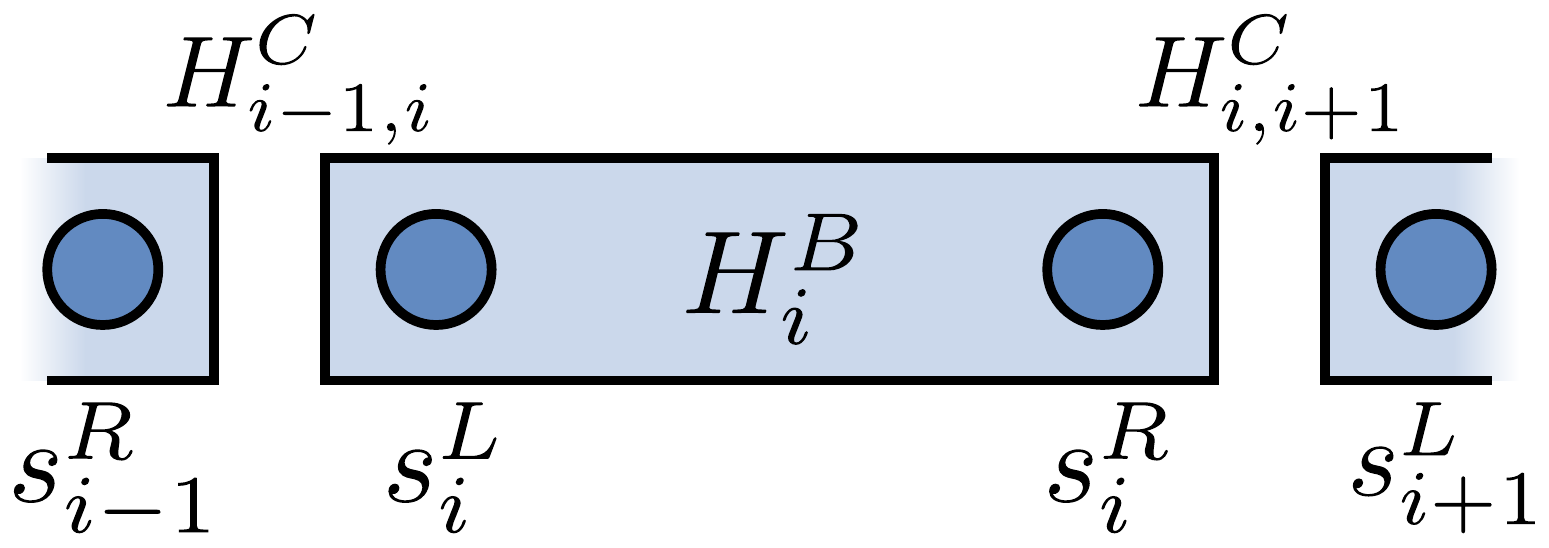}
\caption{(Color online) 
Schematic diagrams of the various SDRG variants. Horizontal lines indicate the 1D spin system.
(a) Traditional MDH SDRG,\cite{MaDH79} spins $\vec{s}_i$, $\vec{s}_{i+1}$ (arrows) with the
  greatest coupling strength, $J_i > J_{k} \forall k\neq i$, are removed and replaced by an effective
  coupling $\tilde{J}$.
(b) SDRG of Westerberg et.\ al.,\cite{WesFSL95} spin pairs are renormalised for the largest energy gap $\Delta_i$ and replaced by an effective spin $\tilde{S}$.
(c) SDRG variant of Hikihara et.\ al.,\cite{HikFS99} the chain
  is decomposed into blocks of spins described by block Hamiltonians $H^{B}$ (shaded rectangles), with left and right spins, respectively, $s^{L}$ and $s^{R}$ (dark dots) on the boundaries of the blocks forming the coupling Hamiltonians, $H^{C}$.}
\label{fig:MDHRSRG}
\label{fig:WestSDRG}
\label{fig:HikiharaRG}
\end{figure}

For \emph{disordered} quantum many-body systems, the strong-disorder renormalization group (SDRG) provides a similarly unifying approach.\cite{IglM05} It was originally devised by Ma, Dasgupta and Hu\cite{MaDH79,DasM80} for the random anti-ferromagnetic (AFM) Heisenberg chain
\begin{equation}
H = \sum_{i=1}^{L-1} J_{i} \vec{s}_{i} \cdot \vec{s}_{i+1},
\label{eq:heisenberg}
\end{equation}
where $\vec{s}_{i}$ is the spin-1/2 operator and $J_{i}$ is the
coupling constant, which takes a random value between $0<J_{i}<J_\text{max}$
according to some probability distribution $P(J)$.
The principle
behind the SDRG is to eliminate the most strongly coupled pairs of
spins and replace them with an effective interaction that couples the
spins at either side of the pair, as shown in Fig.\
\ref{fig:MDHRSRG}a. The pair of spins coupled by $J_\text{max}$ are thought
as being frozen into a singlet ground state as the neighbouring
interactions are significantly weaker --- ultimately leading to the random
singlet phase, which is the ground state of the system.\cite{Fis94,Fis95} 
This freezing of degrees of freedom is remarkably close to an update process in entanglement RG for tensor networks\cite{Vid07} and suggests the possible usefulness of the AdS/CFT correspondence also for disordered spin chains.
By analysing 
the probability of \emph{survival} through the SDRG algorithm it is possible
to predict that mean correlations will have a power-law decay\cite{Fis94} 
with negative power $2$. Similarly, the entanglement entropy can be shown to
scale logarithmically with block size,\cite{RefM04} where 
the amount of entanglement between blocks $A$ and $B$
is quantified by the Von Neumann entropy
\begin{equation}
S_{\text{A}|\text{B}} = -\text{Tr} \rho_{\text{A}} \text{log}_{2} \rho_{\text{A}},
\label{eq:entropy}
\end{equation}
with $\rho_{\text{A}}$ the reduced density matrix obtained by
tracing over the $B$ components of the density matrix.

In this work, we have developed a self-assembling tree tensor network (TTN)
algorithm based on the previous ideas of SDRG.\cite{HikFS99,WesFSL95} This
allows us to calculate properties such as expectation values, correlation
functions and entanglement 
entropy directly and efficiently from the geometry of the TTN. In particular, we find that the
distance dependence of the spin-spin correlation function can be studied not
only via direct calculation of the correlation functions, but also via the
\textit{holographic} distance dependence along the tree network connecting two
sites.
In section \ref{sec:NSDRG} we will briefly review 
the numerical strong disorder renormalization group of Hikihara 
et.\ al.\cite{HikFS99} and define the states and operators that form the basis for our work. 
Section \ref{sec:SDRG_TTN} shows how the numerical
SDRG on a \textit{matrix product operator} (MPO) \emph{self-assembles} the TTN.
Finally, in section \ref{sec:results}  we compute correlation functions and entanglement
entropy (i) directly using the TTN as well as (ii) via simply counting the path lengths and connectivities in the holography. We find that both approaches give consistent results.

\section{The MPO implementation of the SDRG}
\label{sec:NSDRG}

\subsection{The numerical SDRG}

The SDRG method was extended to both
ferromagnetic (FM) and anti-ferromagnetic (AFM) couplings by
Westerberg et.\ al.\cite{WesFSL95,WesFSL97} The approach finds the neighboring
pair of spins $\vec{s}_{i}$, $\vec{s}_{i+1}$  with the greatest energy gap $\Delta_{i}$ between the ground state
and excited state and combines them into a single
effective spin $\tilde{S}$ (Fig.\ \ref{fig:WestSDRG}b). The effective couplings between the
new spin and its neighbours are then recalculated using Clebsch-Gordan coefficients and the new
gaps $\tilde\Delta_{i-1}$ and $\tilde\Delta_{i}$ updated. 
SDRG was once more extended by Hikihara
et.\ al.\cite{HikFS99} to include higher states at each decimation, in
the spirit of the numerical renormalization group (NRG)\cite{Wil75} and the DMRG.\cite{Whi92} 
This method therefore decomposes the system into blocks rather than larger spins allowing for more accurate 
computation of, e.g., the spin-spin correlation functions. The more states that are kept
at each decimation the more accurate the description and is exact in the limit of all states kept.

Consider a point in the algorithm where the Hamiltonian is made up of
blocks $H^{B}_{i}$ at each site and couplings $H^{C}_{i,i+1}$ between
them as in Fig.\ \ref{fig:HikiharaRG}.
The couplings take the form of a two spin Hamiltonian
\begin{equation}
H^{\text{C}}_{i,i+1} = J_{i} \vec{s}^{\text{\,R}}_{i} \cdot \vec{s}^{\text{\,L}}_{i+1},
\end{equation}
where $\vec{s}^{\text{\,R}}_{i}$ is the spin operator of the right
hand spin of block $i$ and $\vec{s}^{\text{\,L}}_{i+1}$ is the left
hand spin of block $i+1$. In full the Hamiltonian is
\begin{equation}
H = \sum_{i=1}^{N_{\text{B}}} H^{\text{B}}_{i} + \sum_{i=1}^{N_{\text{B}}-1} H^{\text{C}}_{i,i+1},
\end{equation}
where $N_{\text{B}}$ is the number of blocks. 

Let us now define the gap $\Delta_{i}$ as the
energy difference between the highest energy SU(2) multiplet that
would be kept and the smallest multiplet that would be discarded in a
renormalization of block $H^{\text{B}}_{i,i+1}$. The scheme works by 
searching for the pair of blocks with the largest
gap $\Delta_{i_{m}}$ and then combines the coupling and the blocks that it
connects into a single block
\begin{equation}
H^{\text{B}}_{i_{m},i_{m}+1} = H^{\text{B}}_{i_{m}} + H^{\text{C}}_{i_{m},i_{m}+1} + H^{\text{B}}_{i_{m}+1}.
\end{equation}
This block and the couplings either side are then renormalized by a 
matrix ($V_{\chi}$) of the eigenvectors corresponding to the lowest $\chi$ eigenvalues
of the block, such that only full SU(2) blocks are kept. The process 
is repeated until the system is represented by one block. The details of the 
algorithm are described in appendix \ref{app:Hik_alg}.

\subsection{Numerical SDRG as an MPO process}

\begin{figure}[t]
(a)\includegraphics[width=0.9\columnwidth]{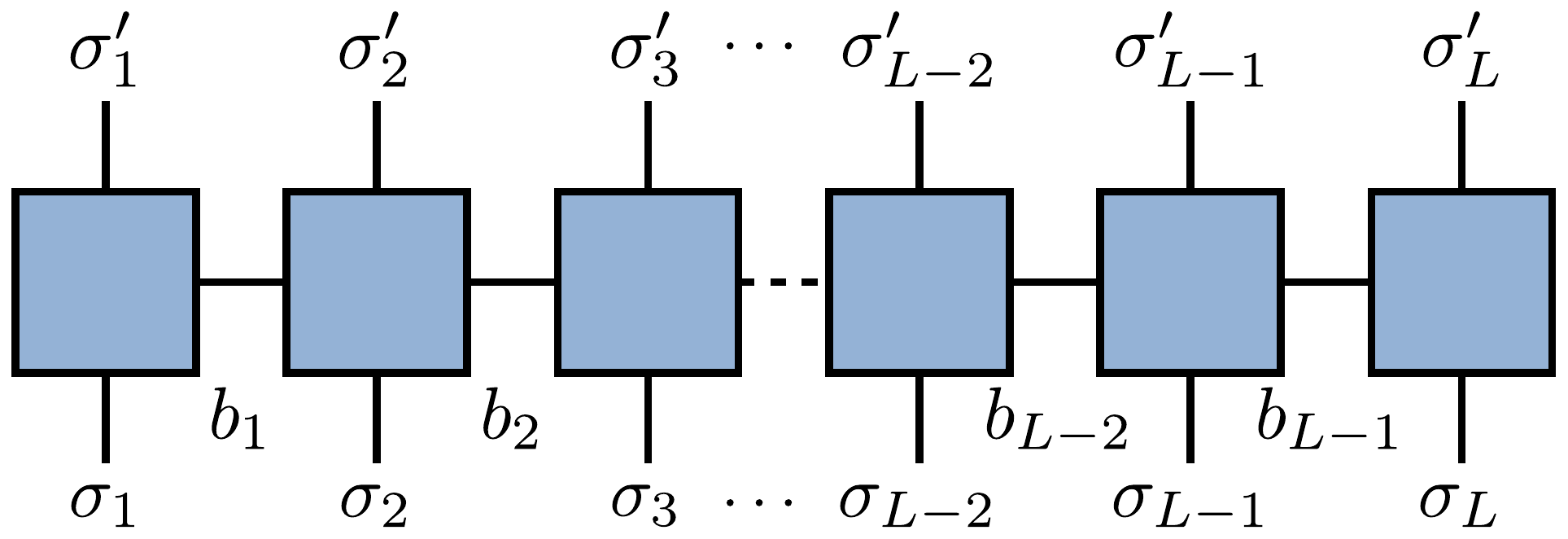}\\\vspace{5pt}
(b)\includegraphics[width=0.9\columnwidth]{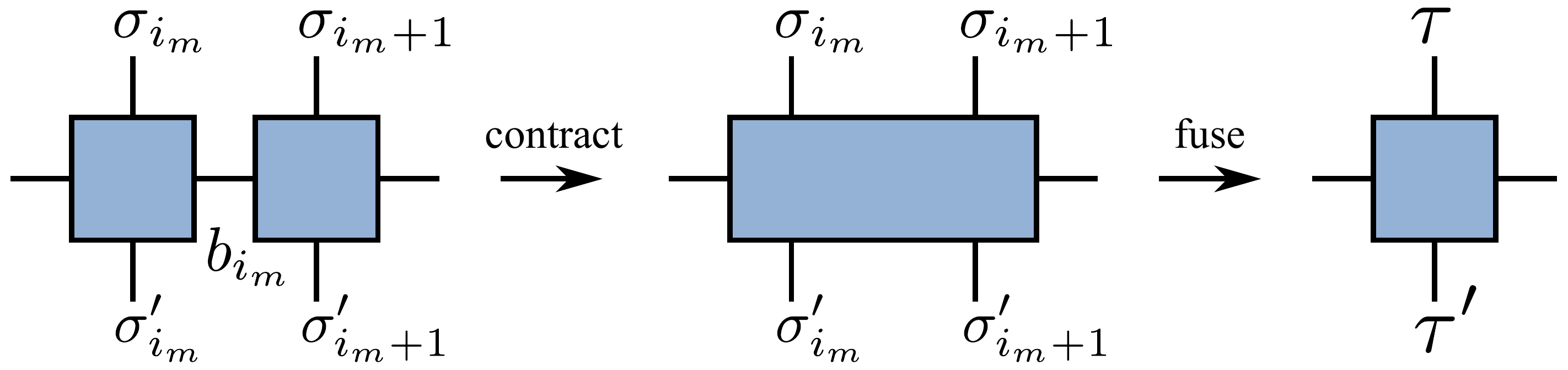}\\\vspace{5pt}
(c)\includegraphics[scale=0.3]{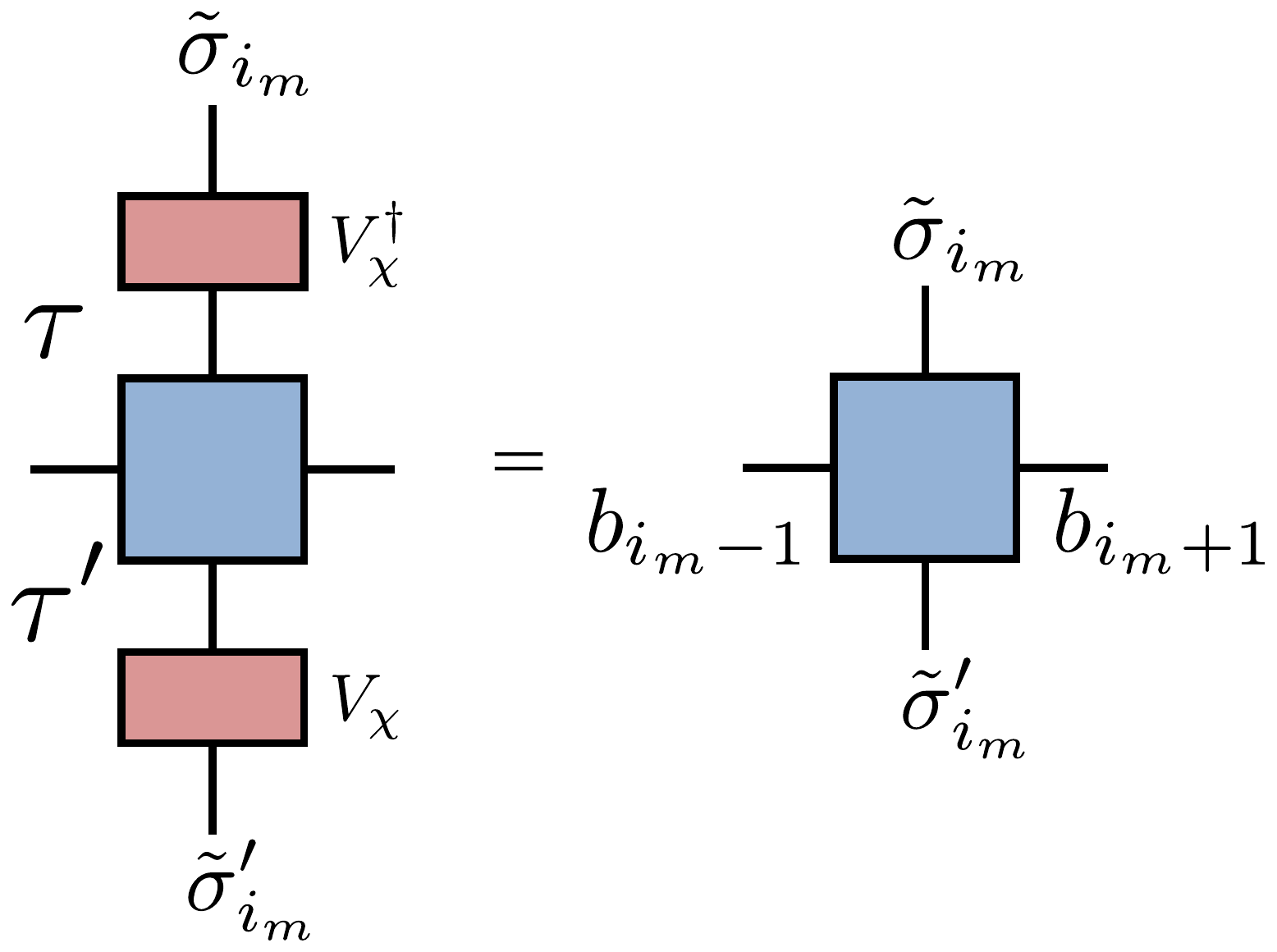}
\caption{(Color online)
(a) Tensor network diagram of the matrix product operator. The (vertical)
  $\sigma$ and $\sigma^{\prime}$ legs denote physical indices
  and couple to the tensor network wavefunction and conjugate. The $b$'s
  are virtual indices (in horizontal direction) and couple the local tensors (blue-shaded squares) of the MPO
  to each other.
(b) The pair of sites with the largest gap $\Delta_{i_{m}}$ is
  found, the MPO tensors for these sites are contracted and the
  physical indices fused to form a matrix.
(c) Contracting the matrices of eigenvectors $V_{\chi}$ (red-shaded rectangle) and
  $V_{\chi}^{\dagger}$ creates a new MPO for a coarse-grained system. }
\label{fig:SDRGtn1}
\label{fig:SDRGtn2}
\end{figure}

Hikihara's numerical SDRG can be naturally described as a set of
operations on an MPO (see Appendix \ref{sec:MPO_SDRG} for more details). 
First, we contract the MPO tensors for the pair of sites with the largest 
gap, sites $i_{m}$ and $i_{m}+1$ (Fig.\  \ref{fig:SDRGtn1}a)
\begin{equation}
W^{[i_{m},i_{m}+1]} = \sum_{b_{i_{m}}} W^{\sigma_{i_{m}}, \sigma^{\prime}_{i_{m}}}_{b_{i_{m}-1},b_{i_{m}}} W^{\sigma_{i_{m+1}}, \sigma^{\prime}_{i_{m}+1}}_{b_{i_{m}}, b_{i_{m}+1}}.
\end{equation}
Here we have $\sigma_{i_{m}}= 1, \ldots, \chi$ for the \emph{physical} indices and, for the Heisenberg model \eqref{eq:heisenberg}, the \emph{virtual} indices are $b_{i_m}=1, \ldots, 5$.
Next, we perform an eigenvalue decomposition on the on-site components of
  the new MPO tensor keeping the eigenvectors of the lowest $\chi$ 
  eigenvalues ($V_{\chi}$)
  \begin{equation}
  \Lambda_{\chi} = V_{\chi}^{\dagger}(H^{B}_{i_{m}} \otimes \openone + J_{i_{m}} \vec{s}_{i_{m}}^{R} \cdot \vec{s}_{i_{m}+1}^{L} + \openone \otimes H^{B}_{i_{m}+1})V_{\chi}.
  \end{equation}
  As with the Hikihara's algorithm, only the $\chi$
  eigenvalues that make up full SU(2) multiplets are used.
Then we contract $V_{\chi}$ and $V_{\chi}^{\dagger}$ with the new MPO 
  tensor to perform the renormalization (Fig. \ref{fig:SDRGtn2}b). 
  For the moment write the two-site combined MPO $W^{[i_m,i_m+1]}$ in terms 
  of an effective site with index $\tau=1, \ldots, \chi, \chi+1, \ldots, \chi^2$, i.e.\ $W^{\tau, \tau^{\prime}}_{b_{i_{m}-1}, b_{i_{m}+1}}$.
  Similarly, we can write the set of eigenvectors as $[V_{\chi}]_\tau^{\tilde{\sigma}_{i_{m}}}$. Then 
  the contraction is explicitly given as 
  \begin{equation}
  W^{\tilde{\sigma}_{i_{m}}, \tilde{\sigma}^{\prime}_{i_{m}}}_{b_{i_{m}-1},b_{i_{m}}} 
  = \sum_{\tau, \tau^{\prime}} [V^{\dagger}_{\chi}]^{\tilde{\sigma}_{i_{m}}}_{\tau} W^{\tau, \tau^{\prime}}_{b_{i_{m}-1}, b_{i_{m}}} [V_{\chi}]^{\tilde{\sigma}_{i_{m}}^{\prime}}_{\tau^{\prime}},
  \label{eq:contraction}
  \end{equation}
  where $\tilde{\sigma}_{i_{m}}= 1, \ldots, \chi$ is the spin index of the renormalised site $i_\text{m}$.
Hence we replace sites $i_{m}$ and $i_{m}+1$ with a single renormalized site and relabel the remaining indices.

\begin{figure}[tb]
(a) \includegraphics[scale=0.3]{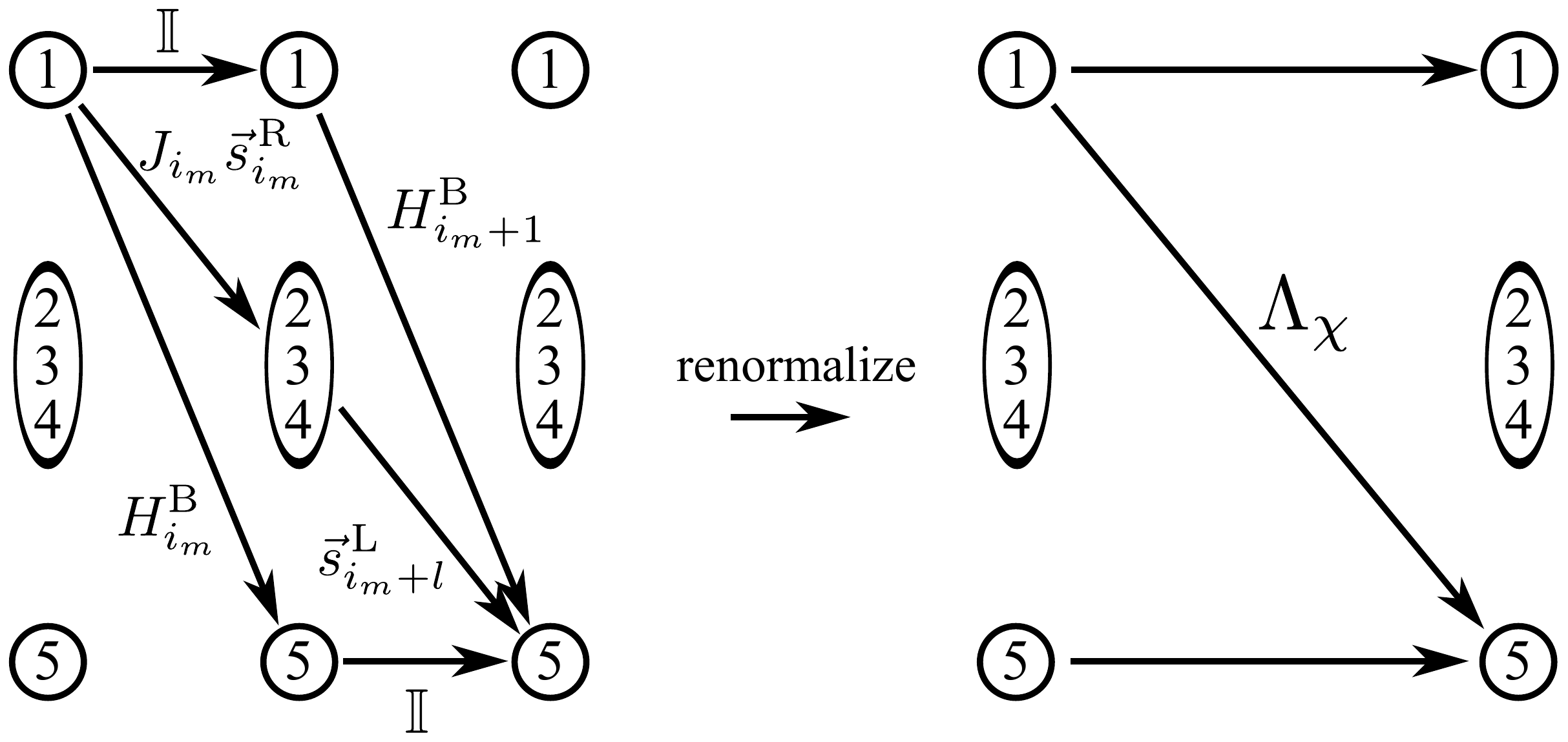}\\
(b) \includegraphics[scale=0.3]{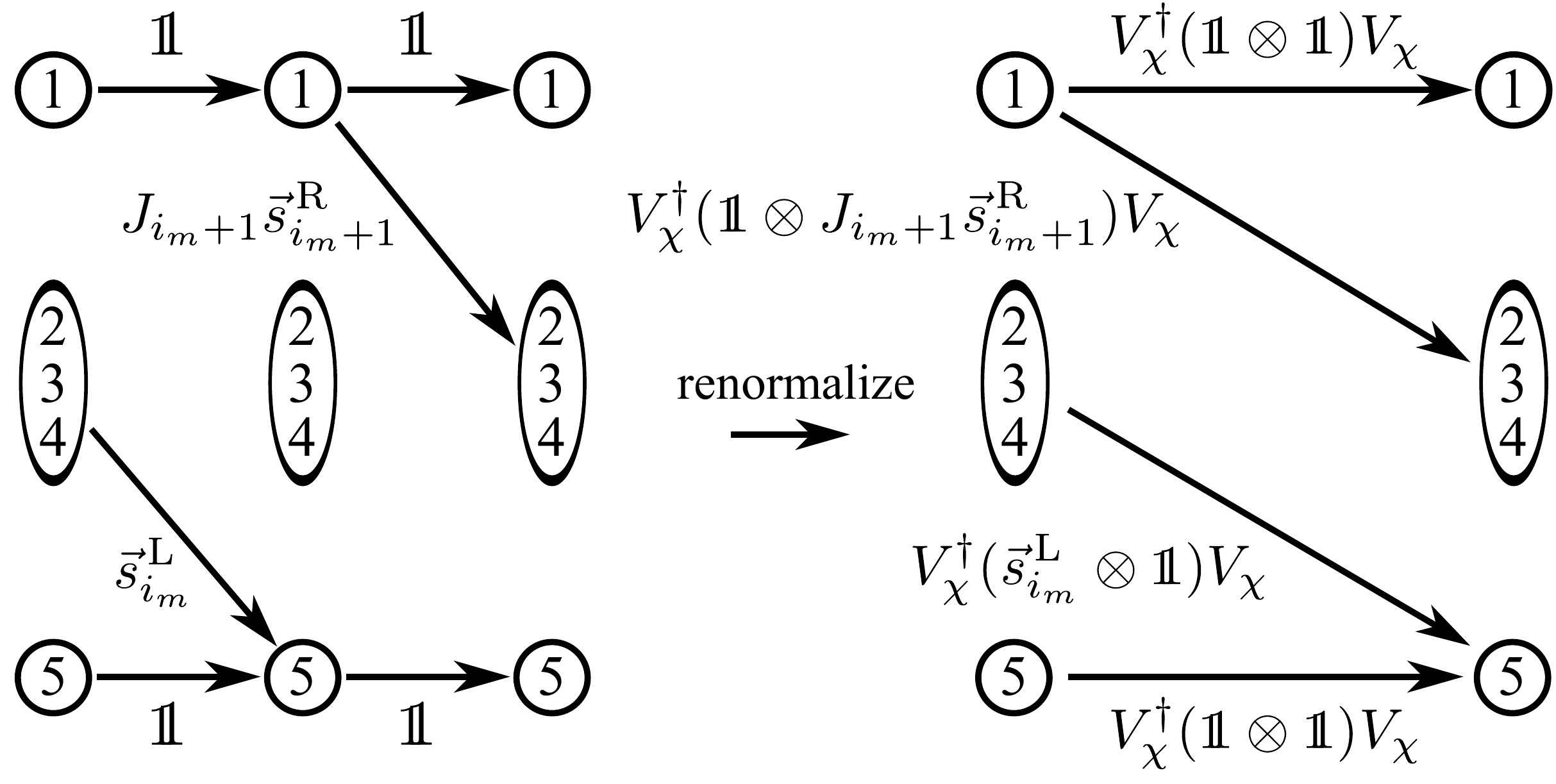}
\caption{
Schematic representation of the contraction step \eqref{eq:contraction} for the Heisenberg Hamiltonian \eqref{eq:heisenberg}. Circles (and ellipses) denote (combined) operator entries in the Heisenberg MPO $W^{[i,i+1]}$, see appendix \ref{sec:mpo} for details.
(a) 
  Renormalizing the on-site components has the effect of
  creating a new on-site component, which is a diagonal matrix of the
  lowest eigenvalues $\Lambda_{\chi}$.
(b)  
  Contracting $V_{\chi}$ and $V_{\chi}^{\dagger}$ has the effect of
  renormalizing the coupling spins in the same way as the Hikihara
  method, storing them as the coupling components of the new MPO
  tensor.}
\label{fig:SDRGMPonsite}
\label{fig:SDRGMPcouplings}
\end{figure}

The contraction makes the on-site component of the new MPO simply a diagonal
  matrix of the lowest $\chi$ eigenvalues $\Lambda_{\chi}$ (Fig.\
  \ref{fig:SDRGMPonsite}a).
It also has the effect of renormalizing the coupling spins just as in
the Hikihara approach (Fig.\ \ref{fig:SDRGMPonsite}b)
  \begin{align}
  \vec{\tilde{s}}^{\,\text{R}}_{i_{m}} =&\: V_{\chi}^{\dagger} ( \openone \otimes \vec{s}^{\,\text{R}}_{i_{m}+1} ) V_{\chi},  \\
  \vec{\tilde{s}}^{\,\text{L}}_{i_{m}} =&\: V_{\chi}^{\dagger} ( \vec{s}^{\,\text{L}}_{i_{m}} \otimes \openone ) V_{\chi}.
  \label{eqn:SDRGcouplings}
  \end{align}
The contraction therefore maps two MPO tensors onto one while preserving the
indexing structure of the MPO.

As the final step, we diagonalize the neighbouring blocks to update the distribution
of gaps. 
The procedure is then repeated until the system is just one site, and we diagonalise to obtain the ground state energy $E_g$ of the system.

\section{Tree tensor networks and SDRG}
\label{sec:SDRG_TTN}

\begin{figure}[tb]
(a) \includegraphics[scale=0.35]{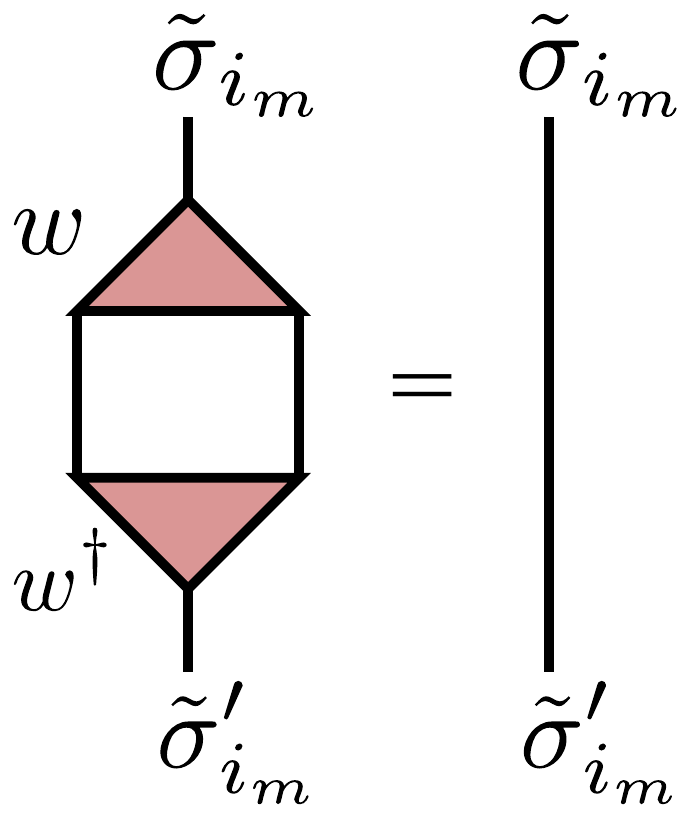}\qquad
(b) \includegraphics[scale=0.25]{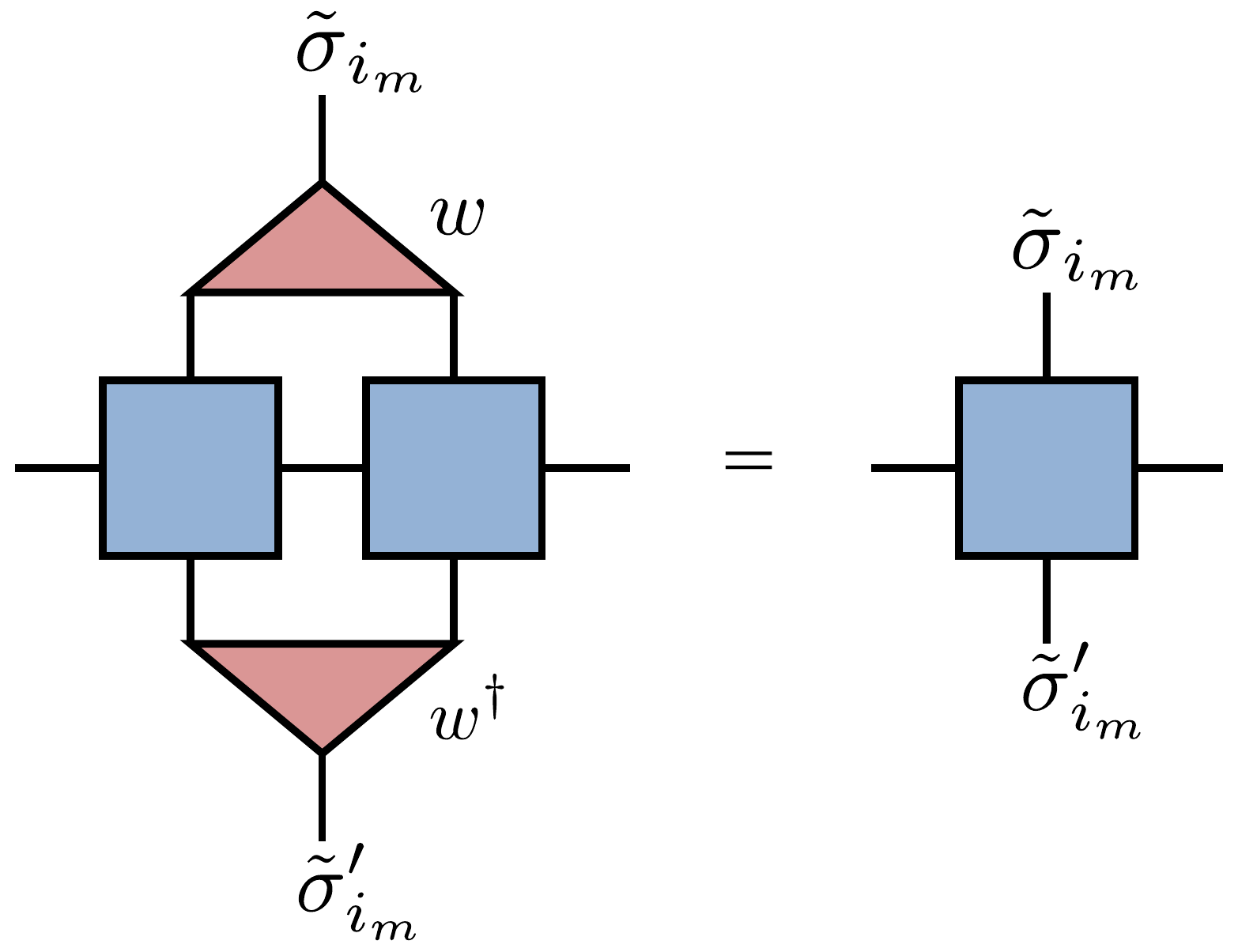}
\caption{(Color online)
(a) Schematic representation of the isometric property $ww^{\dagger}=\openone$ given by 
  Eq.\ (\ref{eqn:isometric}).
(b) One step in the MPO SDRG algorithm in terms of isometric
  tensors $w$. Triangles (red-shaded) denote the isometries, squares are as in Fig.\ \ref{fig:SDRGtn1}.}
\label{fig:isometric}
\label{fig:MPOreno}
\end{figure}

The MPO description of SDRG given above amounts to a coarse-graining
mechanism that acts on the operator. Alternatively, we can view it as a
multi-level tensor network wavefunction acting on the original
operator. To illustrate this, we can \emph{split} the $\tau$
index of $V_{\chi}^{\dagger}$ as in Eq.\ \eqref{eq:contraction} back to the original spin indices 
$\sigma_{i_{m}}$, $\sigma_{i_{m}+1}$ to create an \emph{isometric} tensor 
or \emph{isometry} $\left[V^{\dagger}_{\chi}\right]_{\tau}^{\tilde{\sigma}_{i_\text{m}}}\equiv \left[ w \right]^{\tilde{\sigma}_{i_{m}}}_{\sigma_{i_{m}},\sigma_{i_{m}+1}}$.\cite{EveV09} The isometric property means that 
\begin{equation}
\sum_{\sigma_{i_{m}},\sigma_{i_{m}+1}} [w]^{\tilde{\sigma}_{i_{m}}}_{\sigma_{i_{m}},\sigma_{i_{m}+1}} [w^{\dagger}]^{\tilde{\sigma}^{\prime}_{i_{m}}}_{\sigma_{i_{m}},\sigma_{i_{m}+1}} = \delta^{\tilde{\sigma}_{i_{m}},\tilde{\sigma}^{\prime}_{i_{m}}},
\label{eqn:isometric}
\end{equation}
or $ww^{\dagger}=\openone\neq w^{\dagger}w$ (Fig.\ \ref{fig:isometric}a).
A renormalization in the SDRG algorithm as in Fig.\ \ref{fig:SDRGtn2}b+c can then be rephrased graphically as in 
Fig.\ \ref{fig:MPOreno}b. This makes the notion of mapping
two MPO tensors to one immediately explicit.

\begin{figure}[tb]
\includegraphics[width=\columnwidth]{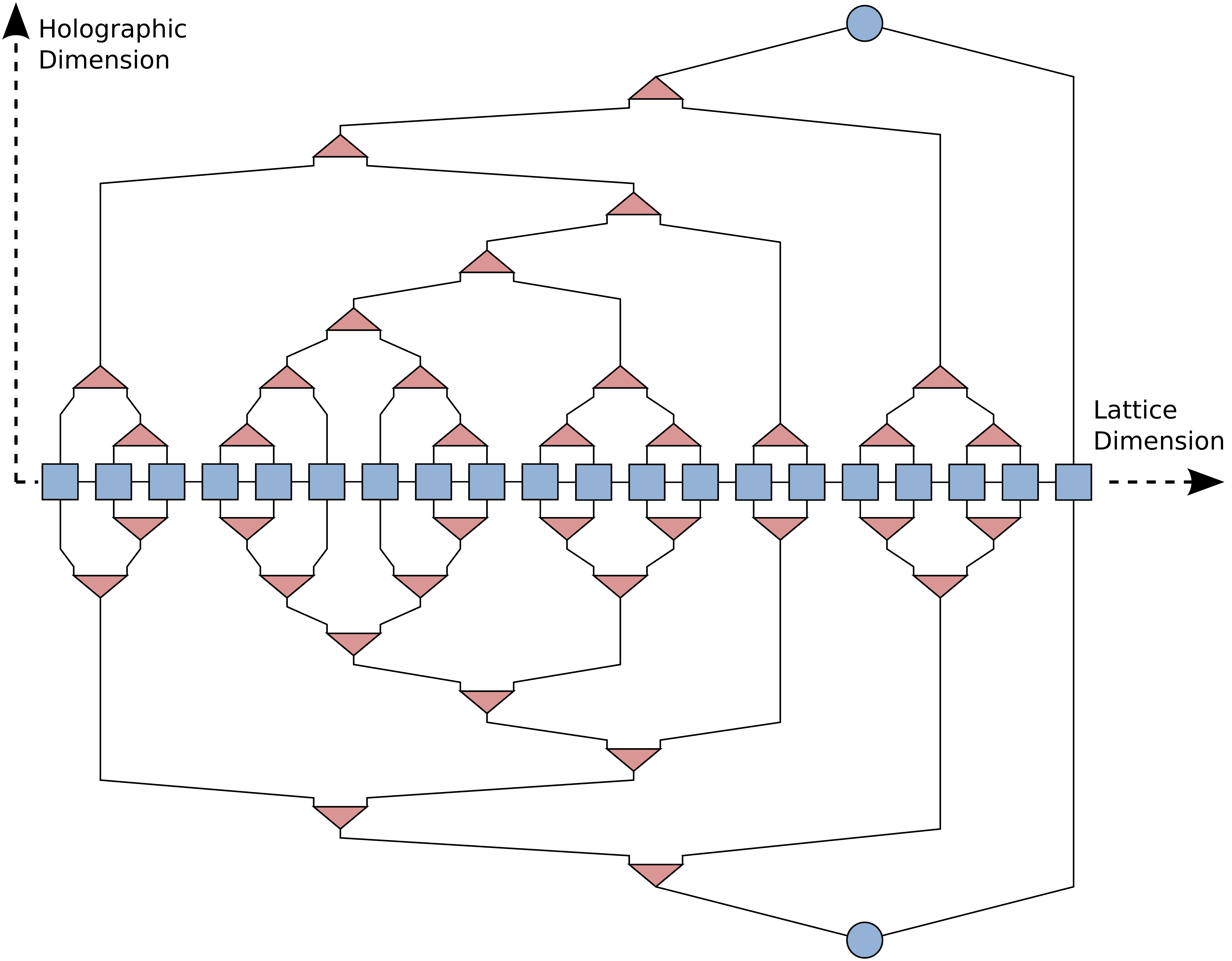}
\caption{(Color online) The SDRG algorithm as a TTN for a
  chain of $L=20$ sites. The squares are the MPOs (i.e.\ the spin 
  operators), triangles are isometric tensors and solid lines denote summations over physical (vertical) and virtual (horizontal) indices as before. The
  circle indicates the \emph{top tensor}, i.e.\ the ground state
  eigenvector of the coarse-grained system. Lattice and holographic dimensions are indicated by the dashed arrows.}
\label{fig:SDRGTTN}
\end{figure}

\begin{figure*}[bt]
(a)\includegraphics[width=0.5\textwidth]{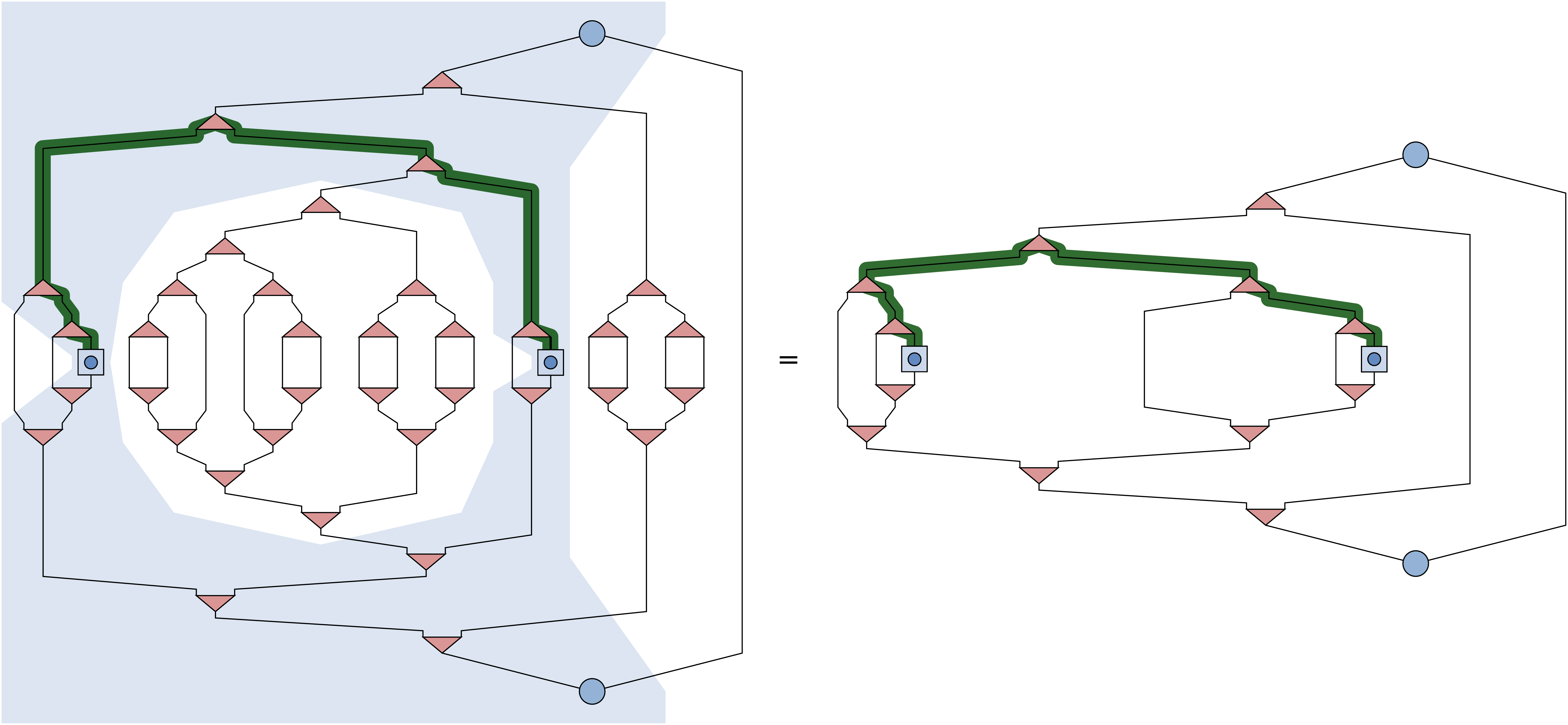}
(b)\includegraphics[width=0.4\textwidth]{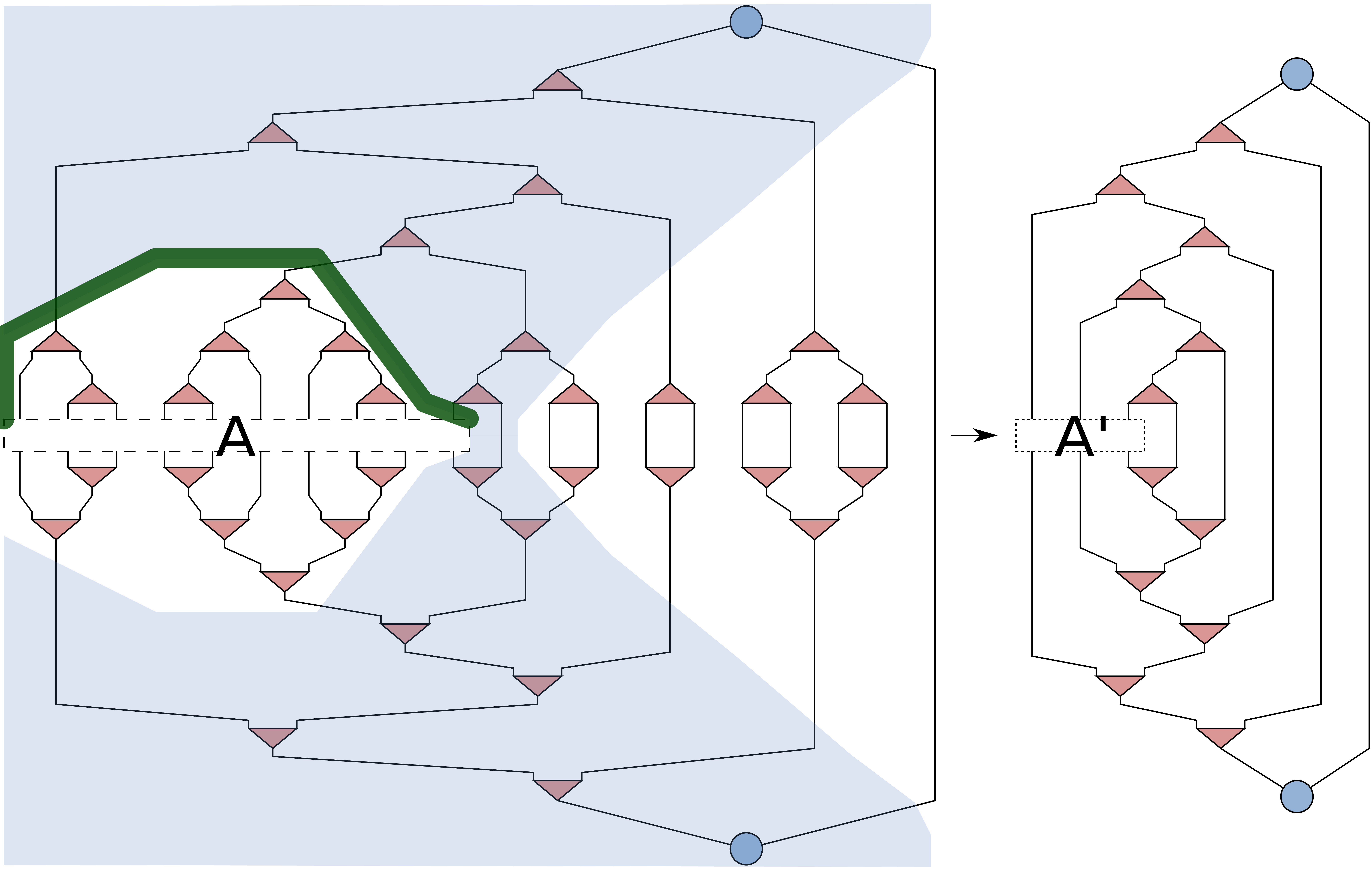}
\caption{(Color online) Diagram showing the TTN form of (a) the correlation function 
  $\langle \vec{s}_{3} \cdot \vec{s}_{15} \rangle$ and (b) the reduced density matrix $\rho_\text{A}$ (for the block $A$ indicated by the dashed rectangle $10$ sites long) of the
  $20$ site system from Fig.\ \ref{fig:SDRGTTN}. Lines and symbols as in Fig.\ \ref{fig:SDRGTTN}. 
  The causal cone in both panels is indicated by a light blue shaded region.
  The bold line in panel (a) shows the path length through the TTN connecting the two sites, whereas the bold line in (b) shows the minimal surface in the TTN between regions $A$ and $B$ (the rest of the chain).
  The diagram in the right-hand side of (b) has been reduced in the horizontal direction to highlight the reduction in complexity due to the isometries.}
\label{fig:SDRGTTNcf}
\label{fig:SDRGTTNdm}
\end{figure*}

When viewed in terms of isometries, the algorithm can be seen to self-assemble 
a tensor network based on the positions of largest gaps before each renormalisation. When written in full, 
it builds an inhomogeneous binary tree tensor network (TTN) as shown in Fig. \ref{fig:SDRGTTN}. 
Tree tensor networks are one of the major areas of tensor network
research and TTNs with regular structures have been extensively
studied.\cite{ShiDV06,TagEV09,CirV09,VerMC08} The isometric nature of
the isometries allows for calculations to be performed in a highly
efficient manner.\cite{TagEV09,Vid08,EveV09} When calculating
expectation values, such as the two-point correlation function (Fig.\
\ref{fig:SDRGTTNcf}a), only those tensors that effect the sites that the
operators act on need to be included, this is known as the
\emph{past causal cone}\cite{EveV09} and is drawn as a blue shadow in the holographic bulk. This allows for a reduction in the number
of contractions that need to be performed to obtain a result.
Calculation of the entanglement (Von Neumann) entropy \eqref{eq:entropy} can also be made
more efficient as shown in Fig.\ \ref{fig:SDRGTTNdm}b. In addition to the reduction due 
to the isometries, we note that the entanglement entropy is not affected by the isometries 
acting just on $A$.\cite{TagEV09} However, the entries in the density matrix will change so we label it $A^{\prime}$.
 
\section{Results}
\label{sec:results}

In the following, we shall compare results for the disordered anti-ferromagnetic Heisenberg model \eqref{eq:heisenberg} when using a modern DMRG implementation, e.g.\ variational MPS (vMPS), with those obtained from our TTN SDRG strategy ({tSDRG}).
The set of couplings ($J_{i}$) shall always be taken from a \emph{box-type} distribution,\cite{HikFS99} i.e.\ constant in the range $0<1-\Delta J/2 < J_{i} < 1+\Delta J/2<2$ and zero outside. Unless stated otherwise, we use strong disorder $\Delta J= 2^{-}$ in the following. We assume open (hard wall) boundary conditions throughout.

\subsection{Convergence and ground-state energies}
\label{sec:energy}

\begin{figure}[tb]
\includegraphics[width=\columnwidth]{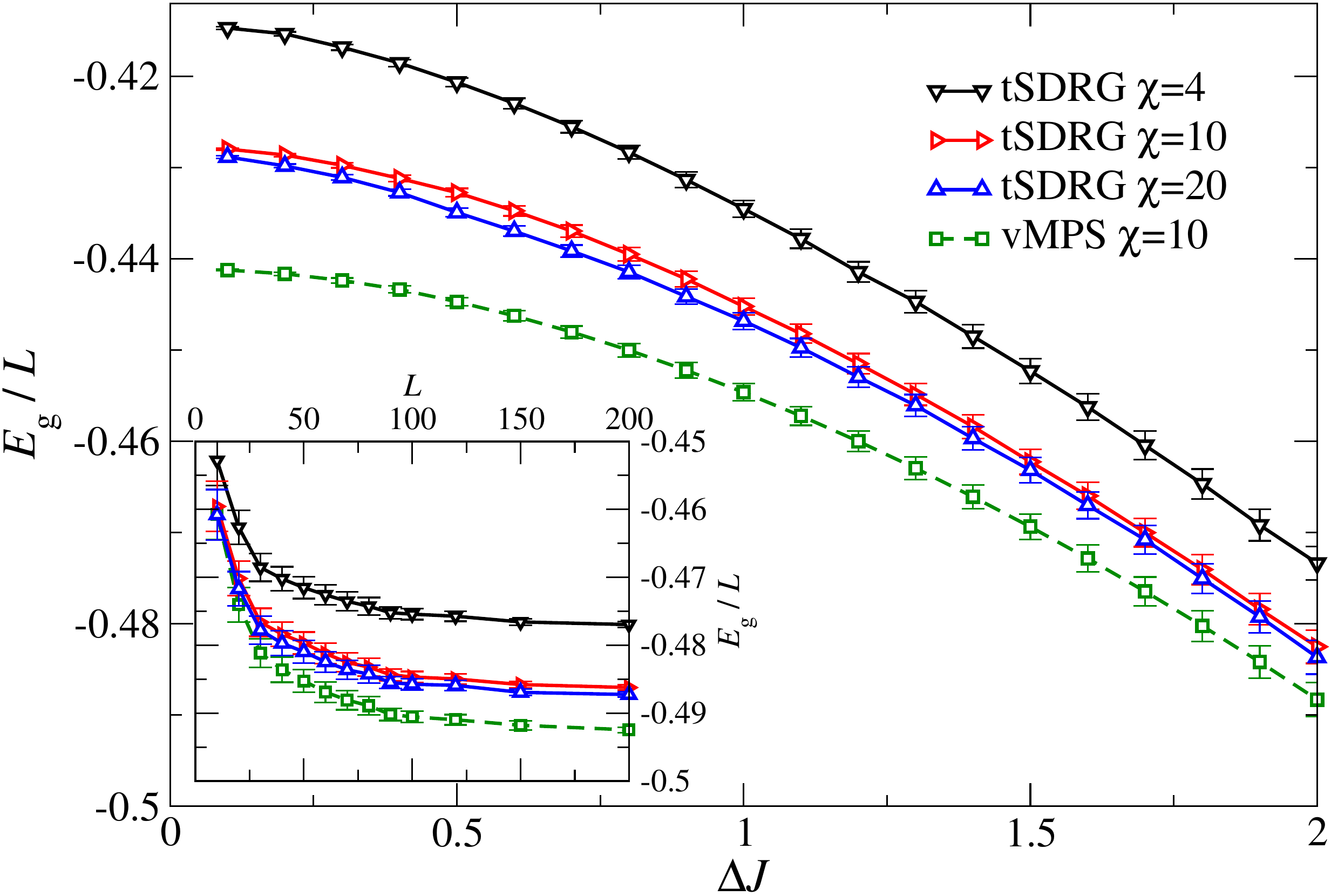}
\caption{(Color online) Ground state energy per site $E_{g}/L$ as a function of disorder 
  $\Delta J$ for system size $L=100$ for tSDRG (solid lines) and variational MPS (dashed). The 
  error bars correspond to the standard error on the mean obtained from averaging over $200$ different disorder configurations and various values of $\chi$. Lines are guides to the eye.
  Inset: System size dependence of $E_{g}/L$ for $\Delta J=2^{-}$. Sizes $L=10$--$80$ have been averaged over
  $500$ disorder configuration, $90$, $100$ and $120$ over $1000$, $150$ and $200$ over $2000$
  configurations, respectively.}
\label{fig:L=100_energy_dis}
\label{fig:L=10-200_energyps}
\end{figure}

In Fig.\ \ref{fig:L=100_energy_dis} (main), we show show the dependence of the disorder averaged ground state energy per site, $E_g/L$, on $\Delta J$ for constant $L$. 
We find that for both vMPS and tSDRG, the  $E_g/L$ values decrease for increasing $\Delta J$, i.e.\ the ground state energy lowers as disorder in the $J_i$ couplings allows the system to form particularly energetically favourable spin configurations.
We also see that the vMPS for the chosen values of $\chi$ and $L$ reaches lower energies. This suggests that it is yet more efficient in finding an approximation to the true ground state energy. However, upon increasing $\Delta J$, the difference between vMPS and tSDRG is getting smaller. This is expected since SDRG is based on the idea that the
contribution from the non singlet interactions is small, which is more accurate an assumption the greater the disorder.
The figure also shows that increasing $\chi$ can considerably improve the results of the tSDRG.\footnote{Let us emphasise that we expect a \emph{variational} tSDRG to be at least as good as our vMPS. Here, however, we concentrate predominately on showing the validity and usefulness of a TTN approach to disordered chains. In order to retain clarity, we thus refrain from increasing algorithmic complexity.}

In Fig.\ \ref{fig:L=10-200_energyps} (inset) we show $E_g/L$ as a function of $L$ for various values of $\chi$ at the strongest permissible disorder $\Delta J=2^{-}$. We find that the values of $E_g/L$ have do not vary much anymore for system sizes $L \geq 100$. Conversely, $E_g/L$ values for $L<100$ are clearly dominated by the presence of the open boundary conditions.

\subsection{Correlation functions}
\label{sec:correlation}

The correlation functions for a strongly disordered Heisenberg chain are
expected to average out to be a power-law decay\cite{Fis94}
\begin{equation}
\langle \langle \vec{s}_{x_{1}} \cdot \vec{s}_{x_{2}} \rangle \rangle \sim \frac{(-1)^{x_{2}-x_{1}}}{|x_{2}-x_{1}|^{2}},
\label{eq:powerlaw}
\end{equation}
where $\langle \langle \vec{s}_{x_{1}} \cdot \vec{s}_{x_{2}} \rangle
\rangle$ is understood to be the disorder averaged expectation value of the two
point spin-spin correlation function. This $r^{-2}$ scaling of the
correlation is a feature of the disorder in the system\cite{Fis94} and should be contrasted with the well-known power-law dependence of correlation functions\cite{Voi95} in interaction-driven Luttinger liquids.
\footnote{
The asymptotic behavior for the clean Heisenberg XXX model\cite{Aff98} is
 $\protect\langle s_{r} {s}_{0} \protect\rangle$
$\sim {(-1)^{r} \sqrt{\log r}}/{r}$.
The difference in power of the algebraic decay points to the very different origin of Eq.\ (\protect\ref{eq:powerlaw}).
}

\begin{figure}[tb]
\includegraphics[width=\columnwidth]{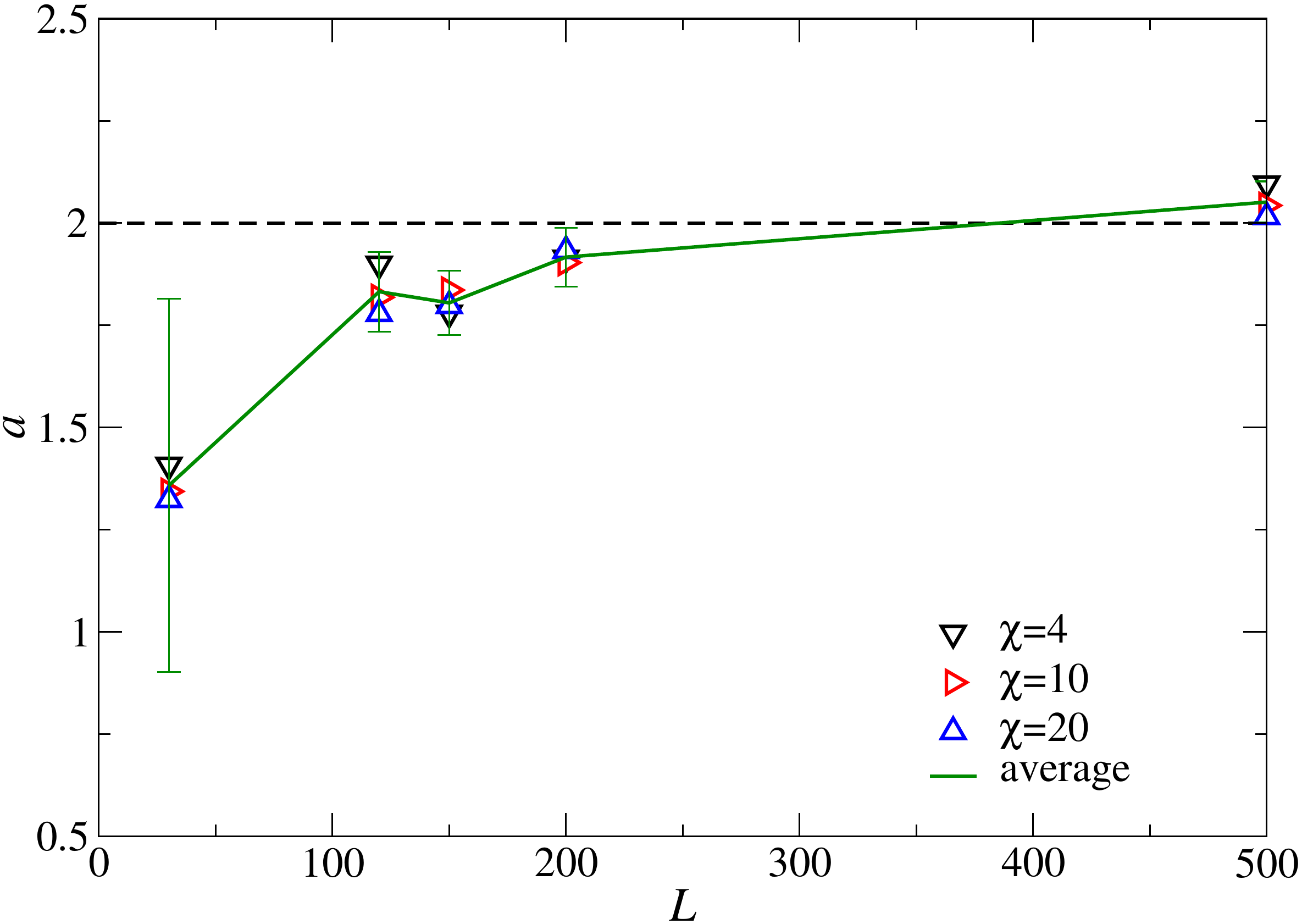}
\caption{(Color online) The scaling parameter $a$ from Eq.\ \eqref{eq:corrSDRG} as a function of system size $L$ for different values of $\chi$ at $\Delta J=2^{-}$. The solid lines are guides to the eye only. The asymptotic value of $a=2$ is indicated by the horizontal dashed line.}
\label{fig:corr_scaling}
\end{figure}

In loop-free tensor networks, correlations scale as $e^{-\alpha D(x_{1},x_{2})}$, where $D(x_{1},x_{2})$
is the number of tensors that connect site $x_{1}$ to $x_{2}$.\cite{EveV11}
DMRG is based on the MPS and as such it has one tensor per site, i.e.\
$D_{\text{MPS}} \approx |x_{2}-x_{1}|$.  Therefore correlations in DMRG
scale exponentially. This suggests that for long chains it
will be necessary to keep large numbers of states to be able to
model a power-law correlation of the system.\cite{EveV11} 
tSDRG on the other hand has a
holographic geometry based on a random TTN with path
length $D_{\text{TTN}} \approx \log |x_{2}-x_{1}|$, i.e.\ scaling logarithmically with distance when
averaged. This makes it much more suited to capture the desired power law decay
\begin{align}
\langle \langle \vec{s}_{x_{1}} \cdot \vec{s}_{x_{2}} \rangle \rangle \sim&\: e^{-\alpha \langle D_{\text{TTN}}(x_{1},x_{2}) \rangle} \nonumber \\
    \sim&\: e^{-\alpha \text{log}|x_{2} - x_{1}|} \sim \: |x_{2} - x_{1}|^{-a}.
\label{eq:corrSDRG}
\end{align}
\begin{figure}[tbh]
\includegraphics[width=\columnwidth]{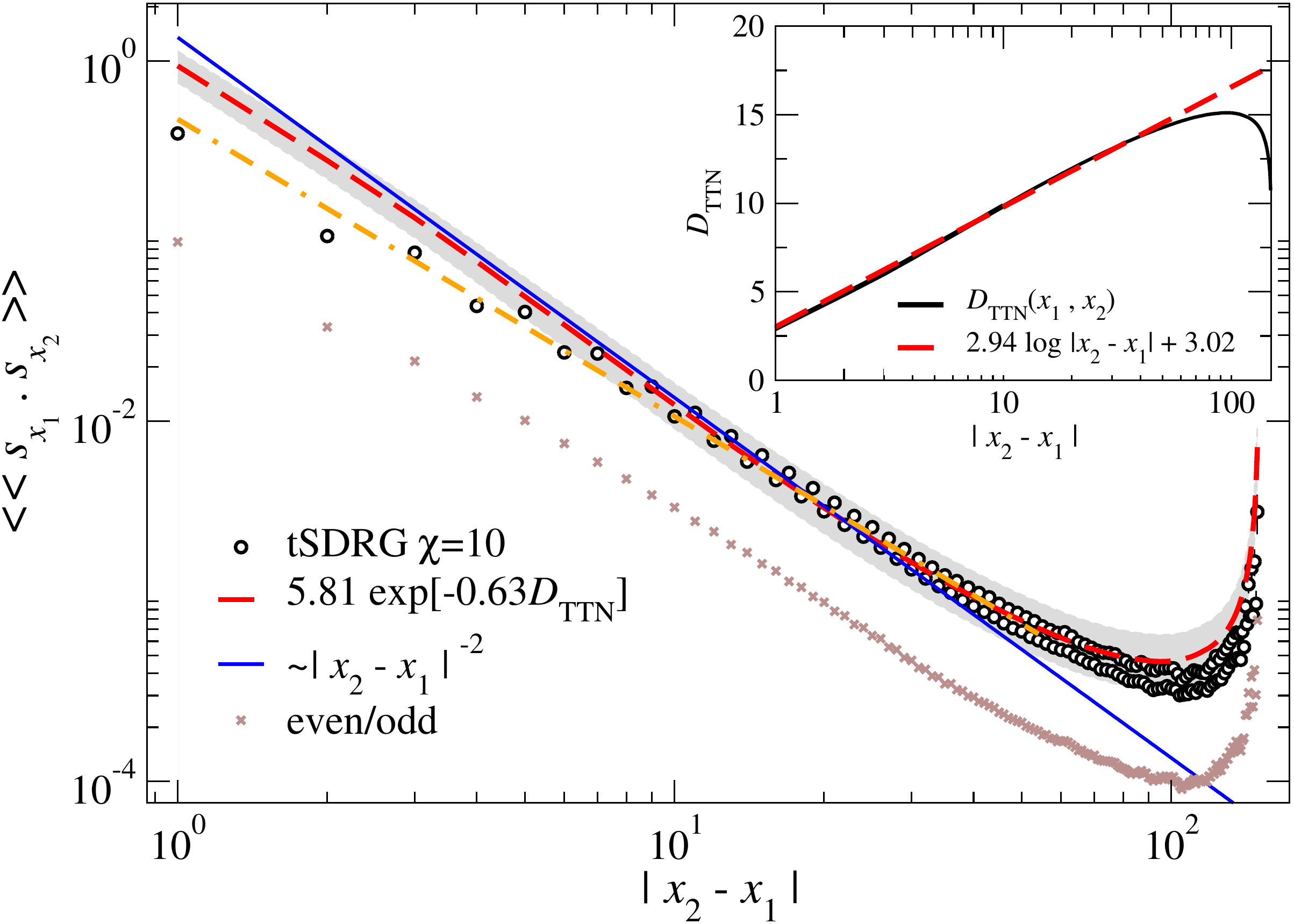}
\caption{(Color online) Correlation function for $L=150$ and $\Delta J=2^{-}$ averaged over
  $2000$ samples for the direct calculation of $\langle \langle \vec{s}_{x_{1}} \cdot \vec{s}_{x_{2}} \rangle \rangle$ (black circles) and also via the holographic approach \eqref{eq:corrSDRG} using $D_{\text{TTN}}$ (dashed red line with error of mean indicated by the grey shading) such that $\langle \langle \vec{s}_{x_{1}} \cdot \vec{s}_{x_{2}} \rangle
  \rangle \approx (5.81 \pm 0.93) \text{exp}[ -(0.62 \pm 0.02) D_{\text{TTN}}]$. 
  The expected thermodynamic scaling $|x_2-x_1|^{-2}$ is also shown (solid blue line) while the dashed orange line denotes a power-law fit up to $|x_2-x_1|=50$ with slope $1.64$. 
  The (brown) crosses show $\langle \langle \vec{s}_{x_{1}} \cdot \vec{s}_{x_{2}} \rangle \rangle /4$ (for clarity) with all values for even distances $|x_{2} - {x_{1}}|$ multiplied by $1.25$.
  Inset: The holographic path length $D_{\text{TTN}}$
  connecting sites $x_{1}$ and $x_{2}$ averaged over the $2000$
  TTNs (black) and a fit in the logarithmic regime (red).}
\label{fig:L=150_corr_dist}
\end{figure}
%
In Fig.\ \ref{fig:L=150_corr_dist}, we show the behaviour of $\langle \langle \vec{s}_{x_{1}} \cdot \vec{s}_{x_{2}} \rangle \rangle$ computed directly as well as its holographic estimate based on \eqref{eq:corrSDRG}. We find that the behaviour for $|x_2-x_1| \gg 1$ and $|x_2-x_1| < L/2$ is indeed very similar for both approaches. The best fit value for $\alpha$ is $0.62 \pm 0.02$ where the error is the standard error.\footnote{The fitting was performed using the \textit{lsqcurvefit} function in MATLAB version 2013a. The function is based on a trust-region algorithm \cite{MorS83} with weights to take into account the accuracy of the data.}
We find that in the indicated distance regime, both measures of $\langle \langle \vec{s}_{x_{1}} \cdot \vec{s}_{x_{2}} \rangle \rangle$ are consistent with the expected $r^{-2}$ behaviour. For $|x_2-x_1| \gtrsim L/2$ we see that the boundaries lead to an upturn on the behaviour of $\langle \langle \vec{s}_{x_{1}} \cdot \vec{s}_{x_{2}} \rangle \rangle$ for both direct and holographic estimates. This upturn is a result of boundary effects and can easily be understood in terms of the holographic TNN: for $|x_2-x_1| \geq L/2$, the average path length in the tree decreases (cp.\ Fig.\ \ref{fig:SDRGTTNdm}). This is also consistent with  periodic systems where we expect correlation functions to be equal for $|x_2-x_1|=r$ and $L-r$.
In the inset of Fig.\ \ref{fig:L=150_corr_dist} we show the distance dependence of $D_{\text{TTN}}$ with $\chi=10$. For $|x_2-x_1|<L/2$, the data can be described by as linear behaviour in $\log |x_2-x_1|$ with slope $2.94\pm 0.02$. Note that this slope along with the value of $\alpha=0.62\pm 0.01$ gives an estimate of power-law exponent $a= 0.62 \times 2.94 = 1.84 \pm 0.04$ for fixed $L=150$.
Figure \ref{fig:corr_scaling} shows that as $L$ increases, the resulting value of the scaling power $a$ also
increases towards the expected value of $2$ for larger systems upon increasing $L$.
We have also checked that the differences between $\chi=10$ and $20$ remain within the error bars and hence we use $\chi=10$ for calculations of $\langle \langle \vec{s}_{x_{1}} \cdot \vec{s}_{x_{2}} \rangle \rangle$ in Fig.\ \ref{fig:L=150_corr_dist}.
We further note that Fig.\ \ref{fig:L=150_corr_dist} shows a clear difference in the correlation function between even and odd distances. The difference in magnitude is found to be $1/4$ as predicted previously.\cite{HoyVLM07}

\begin{figure}[tb]
\includegraphics[width=\columnwidth]{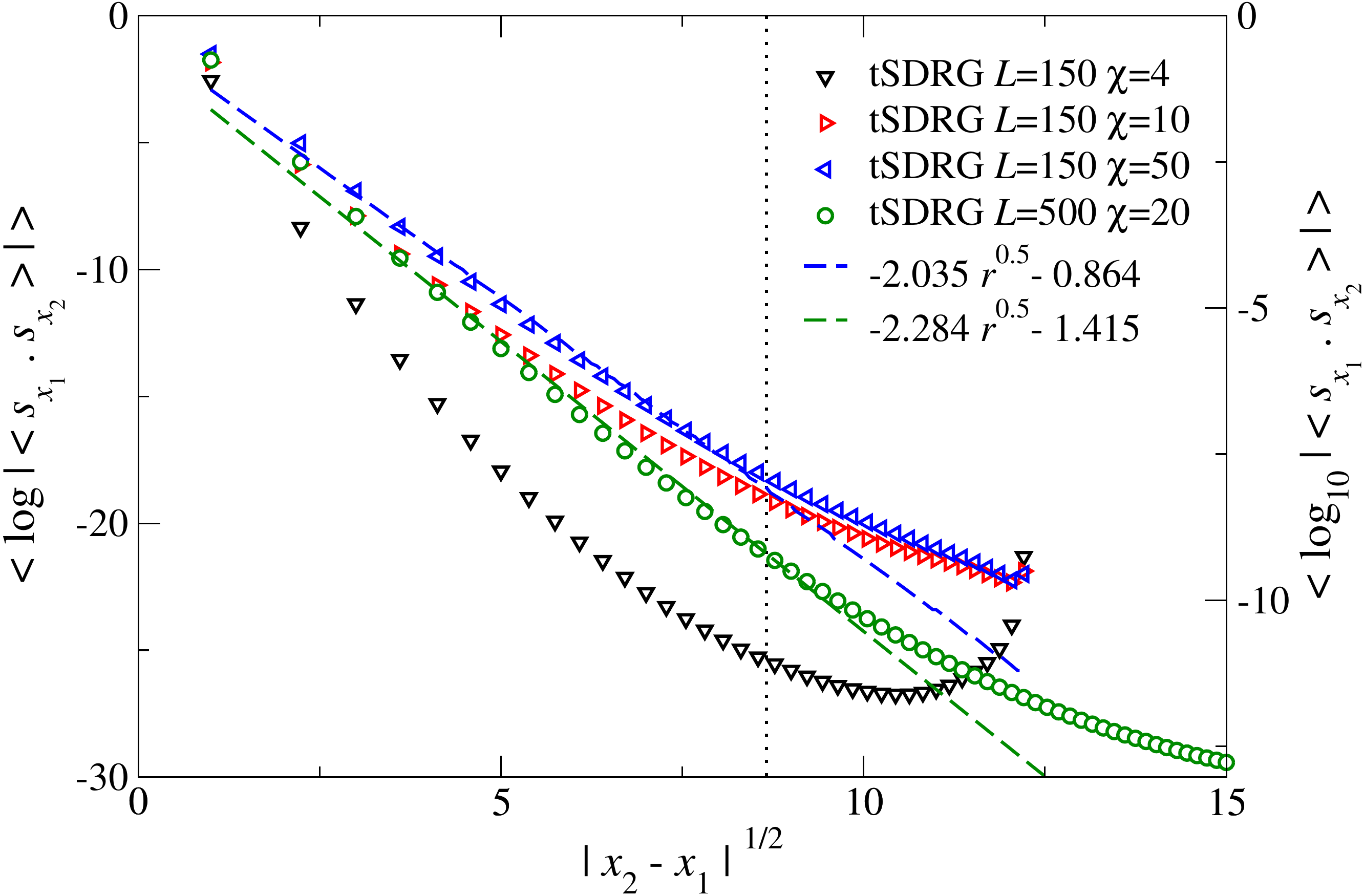}
\caption{(Color online) The \textit{typical} spin correlation function averaged over 2000 samples for $L = 500$ (green circles) and $L = 150$ (triangles) and $\chi$ values as given in the legend. Error bars are within symbol size throughout. The dashed lines are fits to the linear regimes for $L = 150$, $\chi = 50$ (blue) and $L = 500$, $\chi = 20$ (green). The vertical dotted line indicates half the system size for $L = 150$.  }
\label{fig:corr_geo}
\end{figure}

In addition to the power law scaling of mean correlations, it is expected \cite{Fis94} that the \textit{typical} correlations scale as 
\begin{equation}
\langle \text{log} | \langle \vec{s}_{x_{1}} \cdot \vec{s}_{x_{2}} \rangle  | \rangle \sim - | x_{2} - x_{1} |^{1/2},
\label{eq:typmean}
\end{equation}
where the left hand side of \eqref{eq:typmean} is the disorder-averaged mean of the log of the spin correlation function, i.e.\ the log of the geometric mean of the correlation function. Figure \ref{fig:corr_geo} shows that this typical correlation function indeed scales as $| x_{2} - x_{1}|^{1/2}$ and the quality of the fit increases upon increasing $\chi$ and system size. For $L=150$, as $\chi$ is increased from $4$ to $50$, the agreement with \eqref{eq:typmean} improves up to approximately half the system size, at which point boundary effects become important as in Fig.\ \ref{fig:L=150_corr_dist}. 
The typical/geometric mean of the path lengths does not allow to reproduce the typical correlation behaviour \eqref{eq:typmean}, but rather continues to retain a logarithmic scaling behaviour. This suggests that the TTN constructed by our tSDRG selects those path lengths corresponding to mean correlation. Clearly, Eq.\ \eqref{eq:corrSDRG} ignores correlation information stored in the isometry tensors and we expect that its inclusion will recover also the typical correlation behaviour. Indeed, the need to increase $\chi$ in Fig.\ \ref{fig:corr_geo} in order to reproduce \eqref{eq:typmean} already confirms that the tensor content is very important here.

\subsection{Entanglement entropy}
\label{sec:entropy}

In general, the entanglement entropy $S_{\text{A}|\text{B}}$  is difficult to compute as the 
size of the reduced density matrix $\rho_\text{A}$ scales exponentially with the 
size of block $\text{A}$. While for special cases, such as the XX model,\cite{Laf05} $S_{\text{A}|\text{B}}$ can be computed more easily, the general strategy involves finding the eigen- or singular values of $\rho_\text{A}$.\cite{Sch11} 

The TTN representation of tSDRG gives an alternative means of
 calculating $S_{\text{A}|\text{B}}$ for any bipartitions A and B of
the system. In a similar manner to the correlation functions, the geometry of the
tensor network is related to its ability to capture $S_{\text{A}|\text{B}}$. 
Briefly, $S_{\text{A}|\text{B}}$ is proportional to the
minimum number of indices, $n_{A}$, that one would have to cut to separate a
block A of spins from the rest B of the chain (cp.\ Fig.\ \ref{fig:SDRGTTNdm}).\cite{EveV11} This dependence is
related to the famous \emph{area law}, which states that for the ground state of a gapped system, the
entanglement entropy of a region is proportional to the
size of the boundary that separates the two regions.\cite{BomKLS86,EisCP10} The MPS is a
simple line of tensors (cp.\ appendix \ref{sec:MPO_SDRG}) and thus the number of indices that separate
one region from another is a constant and independent of the size of
the block and its position in the chain. Unlike the MPS, for the TTN the
position of the block in the chain alters the number of indices that
have to be cut to separate it from the rest of the system. This suggests that there are spatial regions in
the chain that are more and less entangled, which is likely to be true
for a strongly disordered spin chain. The concept is hence similar to discussing the entanglement in the Ma, Dasgupta, Hu implementation\cite{MaDH79} of SDRG,
where the entanglement entropy is related to the number of singlets
that have to be broken to separate a region from the rest.\cite{RefM04}

\begin{figure}[tb]
\includegraphics[width=\columnwidth]{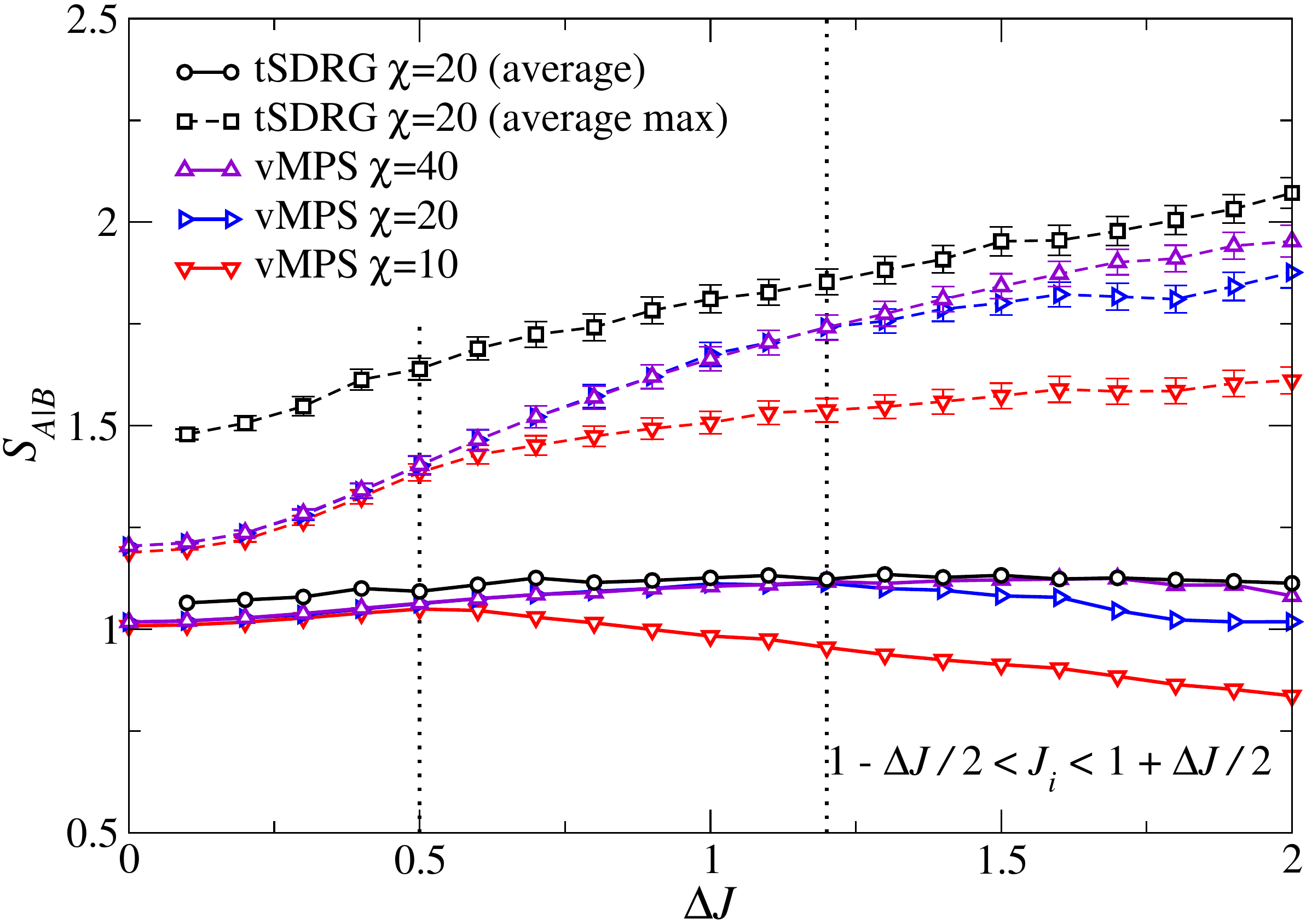}
\caption{(Color online) Entanglement entropy $S_{\text{A}|\text{B}}$ for all possible bipartitions (cp.\ Fig.\ \ref{fig:SDRGTTNdm}) for $L=30$ as a
  function of $\Delta J$ averaged over $100$ disorder configuration using vMPS and tSDRG. 
  Solid lines indicate the arithmetic mean over disorder configurations while dashed lines denote the mean of the maximal $S_{\text{A}|\text{B}}$ values at the chosen $\Delta J$. Lines connecting symbols are guides to the eye only. Error bars denote standard error of the mean when larger than symbol size. The two vertical dotted lines highlight $\Delta J= 0.5$ and $1.2$ as discussed in the text.
  }
\label{fig:DJ_0-2_ee}
\end{figure}
In Fig.\ \ref{fig:DJ_0-2_ee} we show that the average value of $S_{\text{A}|\text{B}}$ remains approximately constant upon increasing the disorder, while the average of the maximal $S_{\text{A}|\text{B}}$ shows a pronounced increase. This indicates that the full distribution of $S_{\text{A}|\text{B}}$ develops long tails with large $S_{\text{A}|\text{B}}$ values when increasing $\Delta J$.
For strong disorders $\Delta J \gtrsim 1.5$ we find that tSDRG is more accurate than vMPS.
The vMPS estimates of $S_{\text{A}|\text{B}}$ are consistently below the values obtained by the tSDRG. Only when increasing $\chi$ do we reduce the deviation. This behaviour is most pronounced for the average of the maximal $S_{\text{A}|\text{B}}$ values. For example, with $\chi=20$, the $S_{\text{A}|\text{B}}$ values obtained for vMPS deviate from the tSDRG results around $\Delta J \approx 1.2$.
Hence we see that an increase in $S_{\text{A}|\text{B}}$ requires a considerable increase in $\chi$ for vMPS to accurately capture the entanglement.
On the other hand, for weak disorders $\Delta J \lesssim 0.5$, vMPS gives consistent results already for small $\chi=10$. The values obtained for  $S_{\text{A}|\text{B}}$ from tSDRG are much higher in this regime. We believe this to be an overestimation of $S_{\text{A}|\text{B}}$ by the tSDRG because, as discussed before, tSDRG selects most strongly the
singlet pairs in the disordered system, which of course become less prevalent for low disorder. 

\begin{figure}[tb]
\includegraphics[width=\columnwidth]{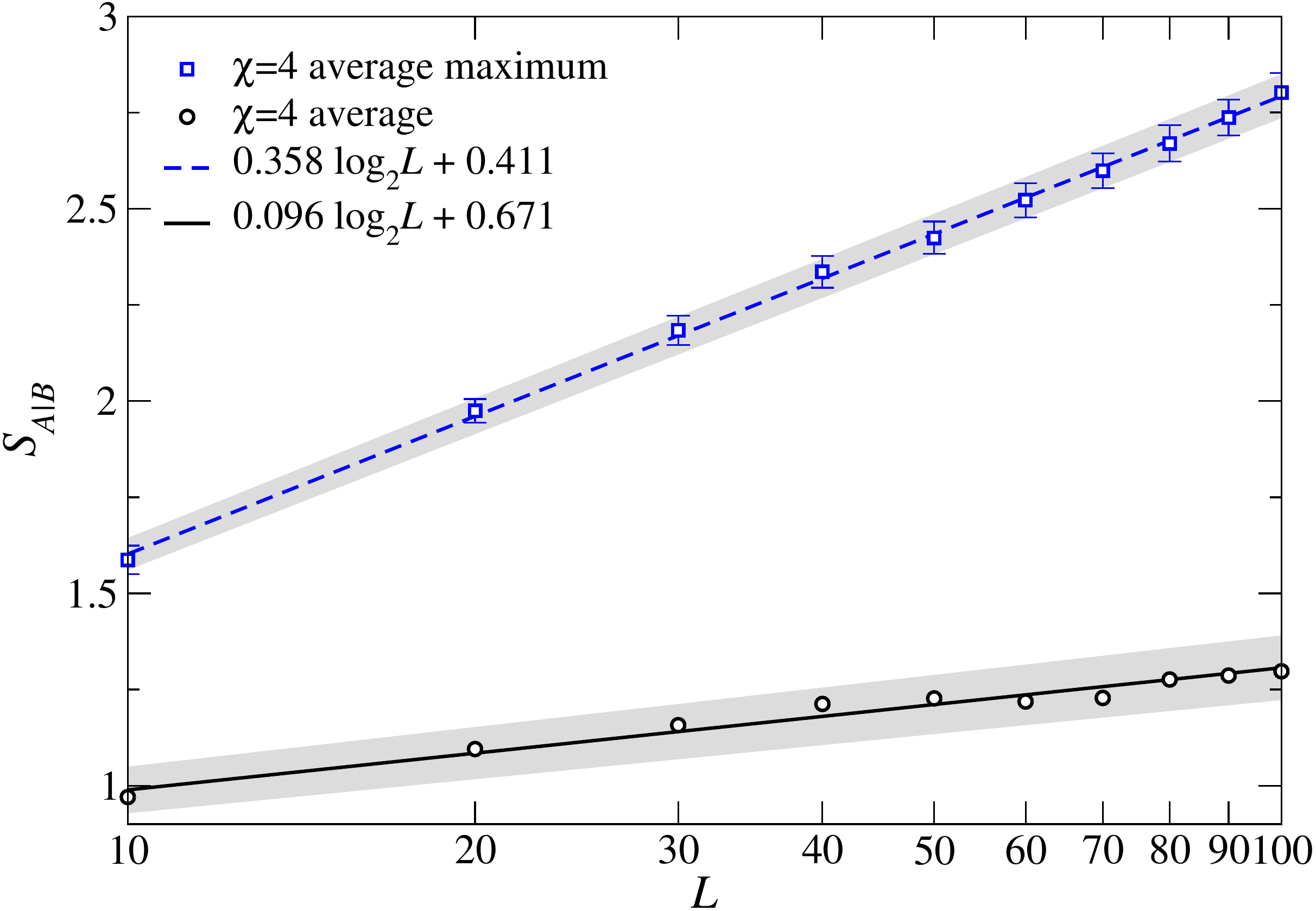}
\caption{(Color online) Entanglement entropy $S_{\text{A}|\text{B}}$ as a
  function of $L$ averaged over $100$ samples and all possible bipartitions (as in Fig.\ \ref{fig:DJ_0-2_ee}) for $\chi=4$ and $\Delta J=2^{-}$. The dashed blue line is the fit $(0.358 \pm 0.005) \text{log}_{2} L + (0.41 \pm 0.03)$, the solid black line is $(0.096 \pm 0.008) \text{log}_{2} L + (0.67 \pm 0.04)$.
  Error bars denote the standard error of the mean for the $S_{\text{A}|\text{B}}$ values when larger than symbol size while grey shaded regions show the standard error of the indicated fits. 
}
\label{fig:L10-100_ee}
\end{figure}

Figure \ref{fig:L10-100_ee} shows that when $L$ is increased
for $\Delta J = 2$, both the average and average peak values of $S_{\text{A}|\text{B}}$  increase
logarithmically in $L$. This again implies that as $L$ is increased, the $\chi$ value for vMPS needs to be increased also to
be able to capture the entanglement. On the other hand, the holographic nature of the TTN
means that the minimal surface in the network increases with system
size and thus describes this entanglement without the need to increase
$\chi$. Although $S_{\text{A}|\text{B}}$ is therefore captured well by the network, contracting $\rho_\text{A}$ for larger $L$ becomes more increasingly difficult, even with the
simplifications suggested in section \ref{sec:MPO_SDRG}, since the size of
the matrices scales as $O(\chi^{n_\text{A}})$. We therefore have to restrict ourselves to smaller $\chi$ and $L$ values than in sections \ref{sec:energy} and \ref{sec:correlation}.

\begin{figure}[tb]
\includegraphics[width=\columnwidth]{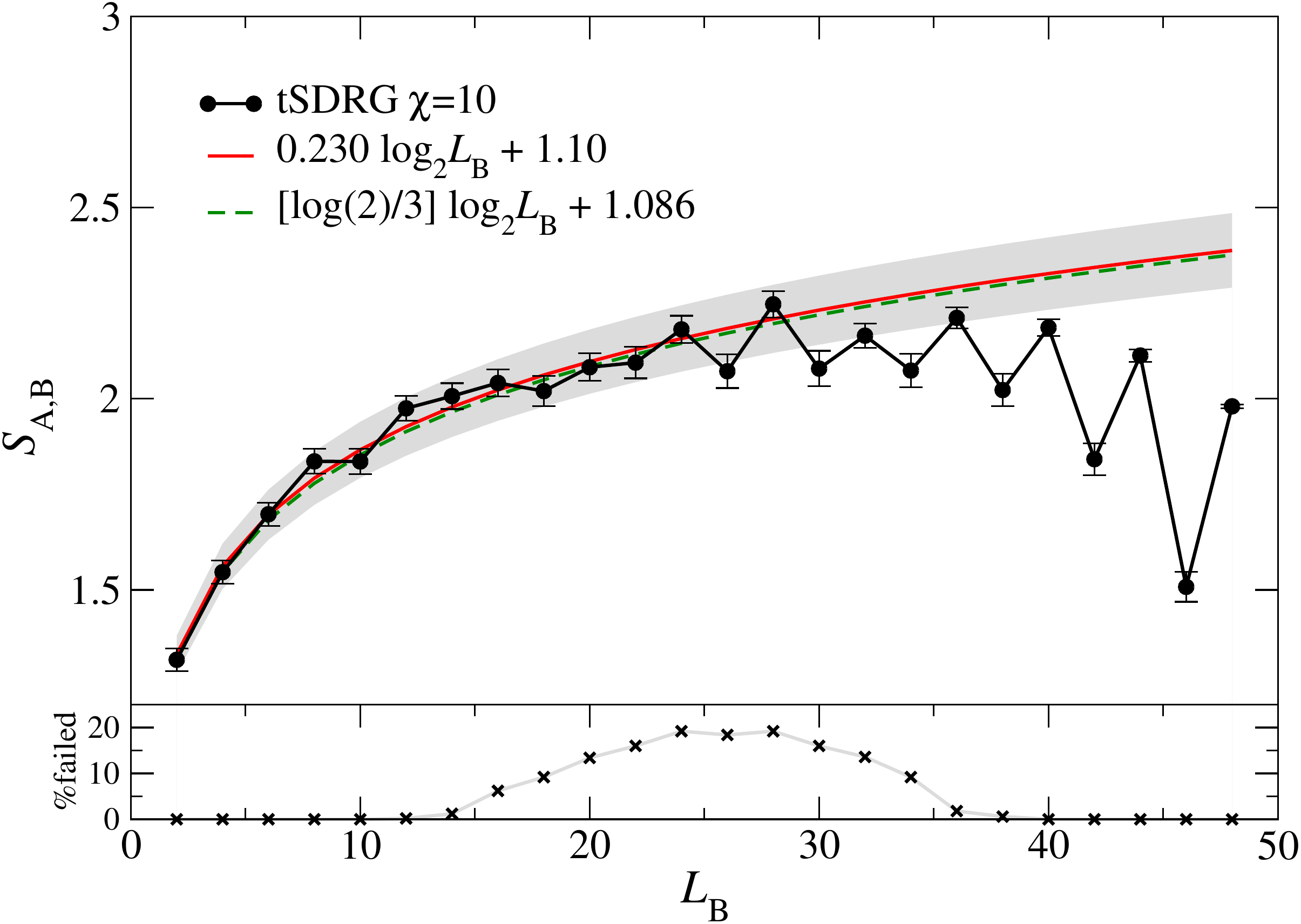}
\caption{(Color online) The entanglement entropy $S_{\text{A},\text{B}}$ (black) averaged over $500$ samples as a
  function of the size of a \emph{block} $L_{\text{B}}$ placed in the middle of a chain with
  $L=50$ for $\chi=10$ and $\Delta J=2^{-}$. The fitting (red, solid line) gives $S_{\text{A}|\text{B}} = (0.22 \pm 0.02) \text{log}_{2} L_{\text{B}} + (1.12 \pm 0.05)$ 
  for $L_{\text{B}} \leq 25$, above which finite size effects dominate. The grey shaded region indicates the accuracy of the fit. The (green) dashed line shows the entanglement scaling \eqref{eq:ee_RefM} from Ref.\ \onlinecite{RefM07} with the vertical position 
  fitted to the point $L_{\text{B}}=2$.  The straight black lines are a guide to the eye only. At the bottom, we show the failure rate in percent (crosses) for different $L_\text{B}$.}
\label{fig:L50_eeb}
\end{figure}

In Refs.\  \onlinecite{RefM04,RefM07}, Refael and 
Moore calculate a \emph{block} entanglement $S_{\text{A}, \text{B}}$ in the random
singlet phase and show that it scales as
\begin{equation}
S_{\text{A}, \text{B}} \sim \frac{\log 2}{3} \log_{2}L_{\text{B}} \approx 0.231\ldots \log_2 L_\text{B},
\label{eq:ee_RefM}
\end{equation}
where region B is a block of extent $L_{\text{B}}$ in the centre of the spin chain. Note that this implies an effective central charge\cite{RefM04} of $\tilde{c}=1\cdot\log 2$.
This is different from the bipartition entanglement $S_{\text{A}|\text{B}}$ that we considered before.
We show the resulting $S_{\text{A}, \text{B}}$ in Figure \ref{fig:L50_eeb}.
The figure clearly indicates that finite size effects become prevalent 
for large $L_\text{B}$, so we fit for $L_{\text{B}} \leq L/2$ only. The resulting scaling behaviour
$S_{\text{A},\text{B}} \approx (0.22 \pm 0.02)\text{log}_{2}L_{\text{B}}$ is consistent with Eq.\ (\ref{eq:ee_RefM}). We note, however, that finite size corrections might still be present at the system size available to us here; ideally one should aim for much larger system sizes.\cite{IglL08}

\begin{figure}[tb]
\includegraphics[width=\columnwidth]{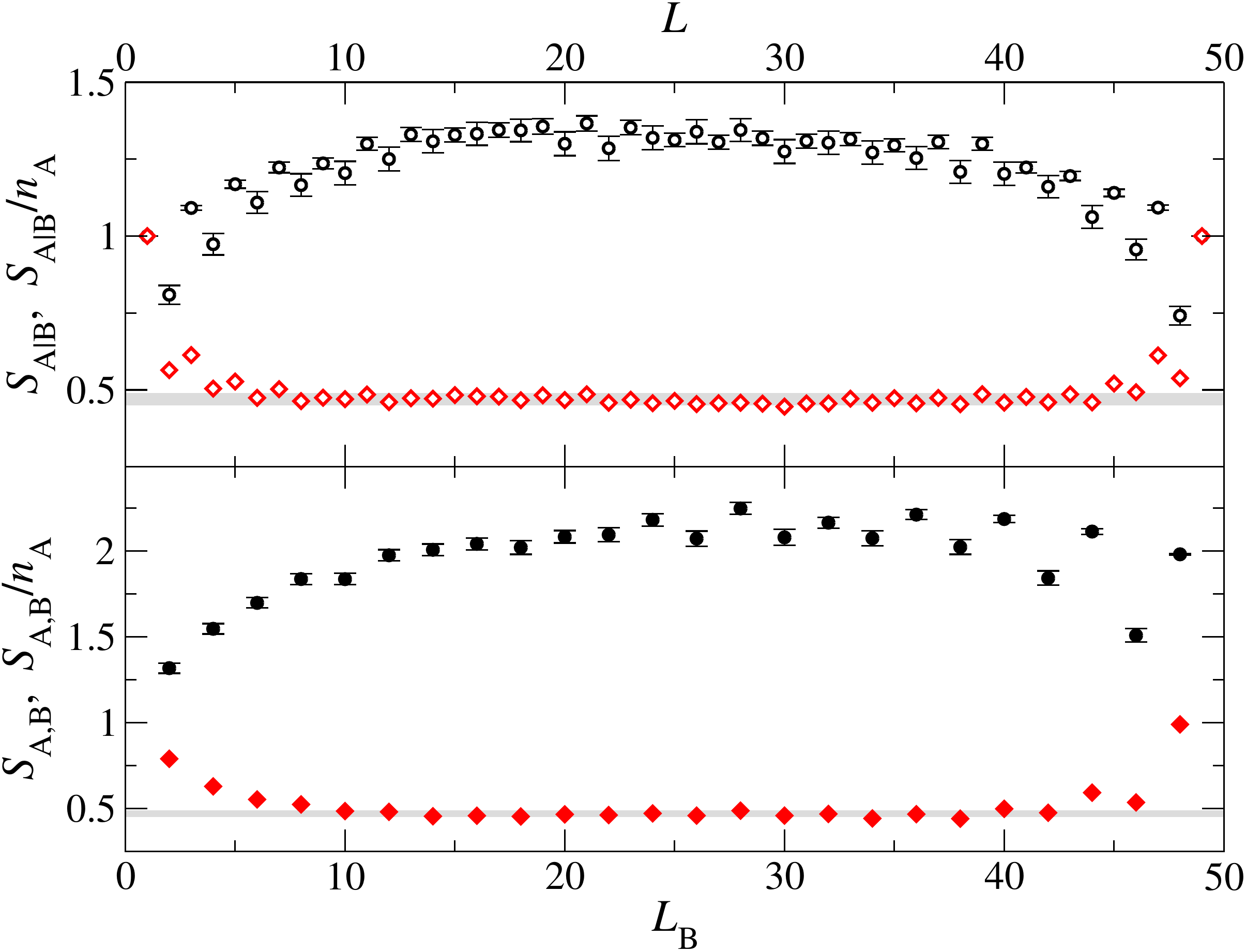}
\caption{(Color online) Entanglement entropy $S$ (black circles) and entanglement entropy per
  bond $S/n_{\text{A}}$ (red diamonds) for bipartitions ${\text{A}|\text{B}}$ (top, open symbols) and blocks ${\text{A}, \text{B}}$ (bottom, filled symbols) with $\chi=10$ and $\Delta J=2^{-}$. The entanglement
  per bond saturates to $0.47 \pm 0.02$ for bipartitions and $0.48 \pm 0.02$ for blocks (grey shaded regions).
}
\label{fig:L50_ee_na}
\end{figure}

We finally also examine the entanglement entropy per bond, $S/n_{\text{A}}$, of a 
TTN for both bipartitions ${\text{A}|\text{B}}$ and blocks $\text{A},\text{B}$ with $\chi=10$
when averaging over $500$ disorder configurations with $L=50$. Figure \ref{fig:L50_ee_na}
shows that away from the boundaries $S/n_{\text{A}}$ saturates to 
the same constant $0.47 \pm 0.02$ for bipartitions and blocks.\footnote{We find $0.42 \pm 0.02$ for $\chi=4$ for both blocks and bipartitions. This might conceivably suggest that $S/n_{\text{A}} = 0.5$ as limiting value for larger $\chi$ and $L$. In turn, this would imply $n_\text{A}= 2\log L_\text{B}/3$. The failure rate for these calculations is $< 1\%$.} This is consistent 
with Ref.\ \onlinecite{EveV11} and implies that the entanglement entropy is proportional to 
the length of the holographic minimal surface that connects the two blocks. 
Note that for $L_\text{B} \sim L/2$, we find that up to $20\%$ of our samples for $\chi=10$ lead to calculations of $S_{\text{A},\text{B}}$ consuming memory beyond $100$GB. This is currently out of reach for us and we disregard the configurations. Nevertheless, we think that this is purely a numerical artefact and does not change the average values of $S_{\text{A},\text{B}}/n_\text{A}$ reported here. Calculations with smaller $\chi$ confirm this.\footnotemark[\value{footnote}]

\section{Conclusion}

In this work, we demonstrate the validity and usefulness of a suitably adaptive tensor network approach to locally disordered one-dimensional quantum many-body systems. In contrast to traditional vMPS approaches to disordered systems, where the initial geometry of the MPS ignores the disorder and only takes it into account at the stage of variational sweeps,\cite{Hal06} our approach incorporates the disorder into the fabric of its tensor network. We believe this strategy to be inherently more suited to disordered systems --- the results presented here show that the accuracy of tSDRG is already comparable to vMPS without including any additional variational updates. This advantage is particularly evident for long-ranged correlations and an entanglement entropy that violates the area law.

Our results furthermore show that, when disorder averaged, a random AFM spin $1/2$  system is well characterised by an effective CFT on the boundary of a discretized holographic bulk. We believe, to the best of our knowledge, that we have thus shown the \emph{quantitative} validity of holography for the first time here. In particular, our spin-spin correlation function, Fig.\ \ref{fig:L=150_corr_dist}, as well as the block and bipartition entanglement entropies, Fig.\ \ref{fig:L50_ee_na} show excellent qualitative and numerical agreement with their holographic counterparts. Such an agreement also reconfirms that the self-assembly of the TTN produces the necessary tensor network geometry.

Whilst here we concentrated on the disordered XXX model, the method should be straightforwardly applicable to the XX and XXZ models as studied by Fisher.\cite{Fis94} Similarly it should work for the Jordan-Wigner transformed equivalent Fermionic models with a disordered hopping parameter.\cite{RamRS14} 
It should also be permissible to implement different forms of disorder, such as aperiodic sequences\cite{JuhZ07} as long as the singlet approximation is valid throughout the renormalization procedure. 
We have checked that tSDRG, just as the SDRG of Hikihara,\cite{HikFS99} is also able to model random FM/AFM couplings that create large effective spins as the renormalization progresses. As such it may be possible to use our approach to study higher spin systems given a suitably high $\chi$. It should also be fairly simple to extend the tSDRG method to periodic systems by introducing a bond between the first and last MPO tensor, which is effectively taking a trace over the MPO. We note that implementation of \textit{on-site disorder}, such as in the random transverse field Ising model,\cite{Fis95} does not appear to have a natural implementation using the local RG outlined in section \ref{sec:NSDRG}. Here it may be possible to implement a tensor network with a different structure, but at the moment it is not clear to us how this would be performed. 

The tensor network approach makes finding other expectation values, i.e.\ in addition to those studied here, straightforward as they are simply the contraction of the set of isometries with a matrix operator. An example is the \textit{string order parameter}\cite{NijR89} that is used to find a hidden topological order in the ground state.\cite{LajCRI05} If the entanglement entropy can be found, so too can the entanglement spectrum, which has become a popular means of characterising many-body wave functions,\cite{LiH08,RegBH09,PolTBO10,LauBSH10,AlbHL12,DenCOM13,LepDS13} for better or for worse.\cite{ChaKS13} Excited states can be found by diagonalising the top tensor and instead of keeping the lowest energy eigenvector, one keeps a suitable set of higher energy eigenvectors. This will only be accurate for low energy excitations as at each step of the renormalization process only the low energy components are kept while information about higher energy modes is discarded. Furthermore, it is possible that when moving far away from the ground state the geometry of the network is no longer appropriate.

Our local RG procedure selects spin pairs based on energy gaps. It is tempting to reformulate this based on the local entanglement content of such pairs. However, it is not straightforward to find such a local measure that captures energies and wave functions well simultaneously. In particular, we do not find a convenient local entanglement measure that would have a simple relation to the local values of $J_i$.
More promising might be the implementation of a variational TTN.\cite{TagEV09} Our initial results suggest that this does indeed improve the energy values, but at considerably increased efforts in implementation and computation --- every disorder configuration of course necessitating its own variationally updated tree structure.


\begin{acknowledgments}
  We are grateful to Andrew Ferris, Glen Evenbly, Jos\'{e} Hoyos and Nick d'Ambrumenil for valuable discussions. We
  would like to thank the EPSRC for financial support (EP/J003476/1) and
  provision of computing resources through the MidPlus Regional HPC
  Centre (EP/K000128/1). AMG would like to thank the organisers and
  participants of the \emph{Networking Tensor Networks 2012}
  workshop at the Centro de Ciencias de Benasque.
\end{acknowledgments}

\appendix
\section{The numerical SDRG Algorithm}
\label{app:Hik_alg}
The algorithm of Ref.\ \onlinecite{HikFS99} can be formulated as follows
\begin{enumerate}
\item Find the coupling Hamiltonian with the largest gap $\Delta_{i_{m}}$
  and create the two-site block.
\item Diagonalize the two-site block to find the $\chi \leq \chi ^{\prime}$ lowest
  eigenvalues ($\Lambda_{\chi}$) and corresponding eigenvectors ($V_{\chi}$)
  such that only full SU(2) multiplets are kept, where $\chi^{\prime}$ is the
  maximum number of eigenvectors and is set at runtime.
\item Set the $\chi$ eigenvalues ($\Lambda_{\chi}$) from the diagonalization
  as the new two-site block, which is equivalent to renormalizing the
  two site block with $V_{\chi}$
\begin{equation}
\tilde{H}^{\text{B}}_{i_{m},i_{m}+1} = V_{\chi}^{\dagger} H^{\text{B}}_{i_{m},i_{m}+1} V_{\chi} = \Lambda_{\chi}.
\end{equation}
\item Renormalize the spin operators on the right and left hand side
  of the new block to update the couplings
\begin{align}
\vec{\tilde{s}}^{\,\text{R}}_{i_{m}} =&\: V_{\chi}^{\dagger} ( \openone \otimes \vec{s}^{\,\text{R}}_{i_{m}+1} ) V_{\chi} \nonumber \\
\vec{\tilde{s}}^{\,\text{L}}_{i_{m}} =&\: V_{\chi}^{\dagger} ( \vec{s}^{\,\text{L}}_{i_{m}} \otimes \openone ) V_{\chi}.
\label{eqn:SDRGnewspins}
\end{align}
\item Diagonalize the neighbouring blocks to get the new gaps.
\item Remove site $i_{m}+1$ and return to step 1.
\end{enumerate}
This process is repeated until the whole system is described by one
block.

\section{Matrix product operators and SDRG}
\label{sec:MPO_SDRG}

\subsection{Matrix product states}

\begin{figure}[tb]
\vspace{0.25cm}
(a)\includegraphics[width=0.9\columnwidth]{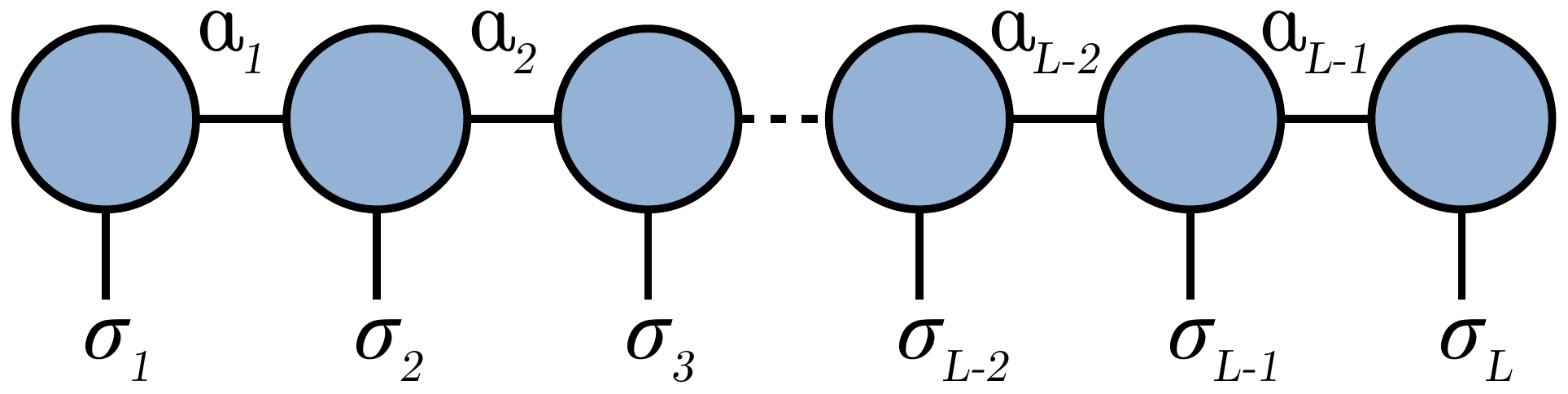}
(b)\includegraphics[width=0.9\columnwidth]{fig-MPOdiag.pdf}
\caption{(Color online) 
(a) Tensor network diagram of the MPS ket. The circles represent the
  $M$ tensors in Eq.\ (\ref{eqn:MPS}), the lines are the indices and
  connected lines represent tensor contractions. The bra state is the same 
  apart from the vertical lines point upwards and the contents of the tensors
  are the complex conjugate.
(b) Tensor network diagram of the matrix product operator. The
  $\sigma$ and $\sigma^{\prime}$ legs are \emph{physical indices}
  and couple to the tensor network wavefunction and conjugate. The $b$
  are \emph{virtual indices} and couple the local tensors of the MPO (squares)
  to each other.}
\label{fig:MPS}
\label{fig:MPOdiag}
\end{figure}

A general wavefunction describing a spin state on a lattice can be written as
\begin{equation}
\ket{\Psi} = \sum_{\sigma_{1}, \dots, \sigma_{L}} C_{\sigma_{1} \dots \sigma_{L}} \ket{\sigma_{1}, \dots, \sigma_{L}},
\label{eqn:TensorNetwork1}
\end{equation}
where $\sigma_{i}$ are the \emph{physical indices} of the
lattice and enumerate the states in the local Hilbert space. The tensor $C_{\sigma_{1} \dots \sigma_{L}}$ can be
decomposed into a \emph{tensor network}, the most common of which is
the \emph{Matrix product state} (MPS)
\begin{equation}
\ket{\Psi} = \sum_{\sigma_{1}, \dots , \sigma_{L}} \sum_{a_{1}, \dots, a_{L-1}} M^{\sigma_{1}}_{a_{1}} M^{\sigma_{2}}_{a_{1} a_{2}} \dots M^{\sigma_{L}}_{a_{L-1}} \ket{\sigma_{1}, \dots, \sigma_{L}}.
\label{eqn:MPS}
\end{equation}
Here, $a_i = 1, \ldots, \chi$ for site $i$ in the bulk.
It is convenient when studying tensor networks, such as the MPS, to
give the equations a diagrammatic form (Fig.\ \ref{fig:MPS}a). Each tensor is
drawn as a shape where each line coming out represents an index and
connected lines represent tensor contractions.

\subsection{Matrix product operators}
\label{sec:mpo}

In a similar manner to the matrix product state, operators acting on
lattice wavefunctions can be decomposed into a network of more simple
tensors. A general operator on a lattice can be written as:
\begin{equation}
\mathcal{O} = \sum_{\sigma_{1}, \ldots, \sigma_{L}} \sum_{\sigma^{\prime}_{1}, \ldots, \sigma^{\prime}_{L}} D_{ \sigma_{1}, \sigma^{\prime}_{1} \dots  \sigma_{L}, \sigma^{\prime}_{L}} \ket{\sigma_{1} \dots \sigma_{L}} \bra{\sigma^{\prime}_{1} \dots \sigma^{\prime}_{L}}.
\label{eqn:operator}
\end{equation}
This can be decomposed into a matrix product form to give a
\emph{matrix product operator} (MPO):
\begin{eqnarray}
\mathcal{O}  &= & \sum_{\sigma_{1}, \ldots, \sigma_{L}} \sum_{\sigma^{\prime}_{1}, \ldots, \sigma^{\prime}_{L}} \sum_{b_{1}, \ldots, b_{L-1}} \times \\ \nonumber
& & W^{\sigma_{1}, \sigma^{\prime}_{1}}_{b_{1}} W^{\sigma_{2}, \sigma^{\prime}_{2}}_{b_{1}, b_{2}} \dots W^{\sigma_{L-1}, \sigma^{\prime}_{L-1}}_{b_{L-2}, b_{L-1}} W^{\sigma_{L}, \sigma^{\prime}_{L}}_{b_{L-1}} \times \\ \nonumber
& & \ket{\sigma_{1} \dots \sigma_{L}} \bra{\sigma^{\prime}_{1} \dots \sigma^{\prime}_{L}},
\label{eqn:MPO}
\end{eqnarray}
where there are two sets of physical indices $\sigma$ and
$\sigma^{\prime}$, which connect to the bra and ket states
respectively. Figure \ref{fig:MPOdiag}b gives the pictorial form of the
MPO.

The Heisenberg Hamiltonian on an open lattice,
\begin{equation}
H_{\text{XXX}} = \sum_{i=1}^{L-1} J_{i} \left[ \frac{1}{2} \left( s^{+}_{i} s^{-}_{i+1} + s^{-}_{i} s^{+}_{i+1-1} \right) + s^{z}_{i} s^{z}_{i+1} \right],
\label{eq:hamiltonian}
\end{equation}
can be encoded as an MPO with
\begin{align}
W_{b_{1}} =&\: \begin{pmatrix} \openone & \frac{J_{1}}{2} s_{1}^{+} & \frac{J_{1}}{2} s_{1}^{-} & J_{1} s_{1}^{z} & 0 \end{pmatrix}, \label{eqn:heisW1} \\
W_{b_{i-1},b_{i}} =&\: \begin{pmatrix} \openone & \frac{J_{i}}{2} s_{i}^{+} & \frac{J_{i}}{2} s_{i}^{-} & J_{i} s_{i}^{z} & 0 \\
0 & 0 & 0 & 0 & s_{i}^{-} \\
0 & 0 & 0 & 0 & s_{i}^{+} \\
0 & 0 & 0 & 0 & s_{i}^{z} \\
0 & 0 & 0 & 0 & \openone \end{pmatrix}, \label{eqn:heisWi} \\
W_{b_{L-1}} =&\: \begin{pmatrix} 0 \\
s_{L}^{-} \\
s_{L}^{+} \\
s_{L}^{z} \\
\openone \end{pmatrix}. \label{eqn:heisWL}
\end{align}
Simply multiplying $W_{b_{1}} W_{b_{1}, b_{2}} \cdots W_{b_{L-2}, b_{L-1}} W_{b_{L-1}}$ results in \eqref{eq:hamiltonian}. 
The top right element of Eq. (\ref{eqn:heisWi}) and equivalent
elements in Eqns. (\ref{eqn:heisW1}) and (\ref{eqn:heisWL}) are referred 
to as the \emph{on-site} elements. This is where an external magnetic field 
of the form $h_{i}S_{i}^{z}$ would be introduced. Furthermore, it is possible to 
include longer range interactions in the elements away from the top and
right row and column.\cite{CroB08,FroND10}

\begin{figure}[tb]
\includegraphics[scale=0.3]{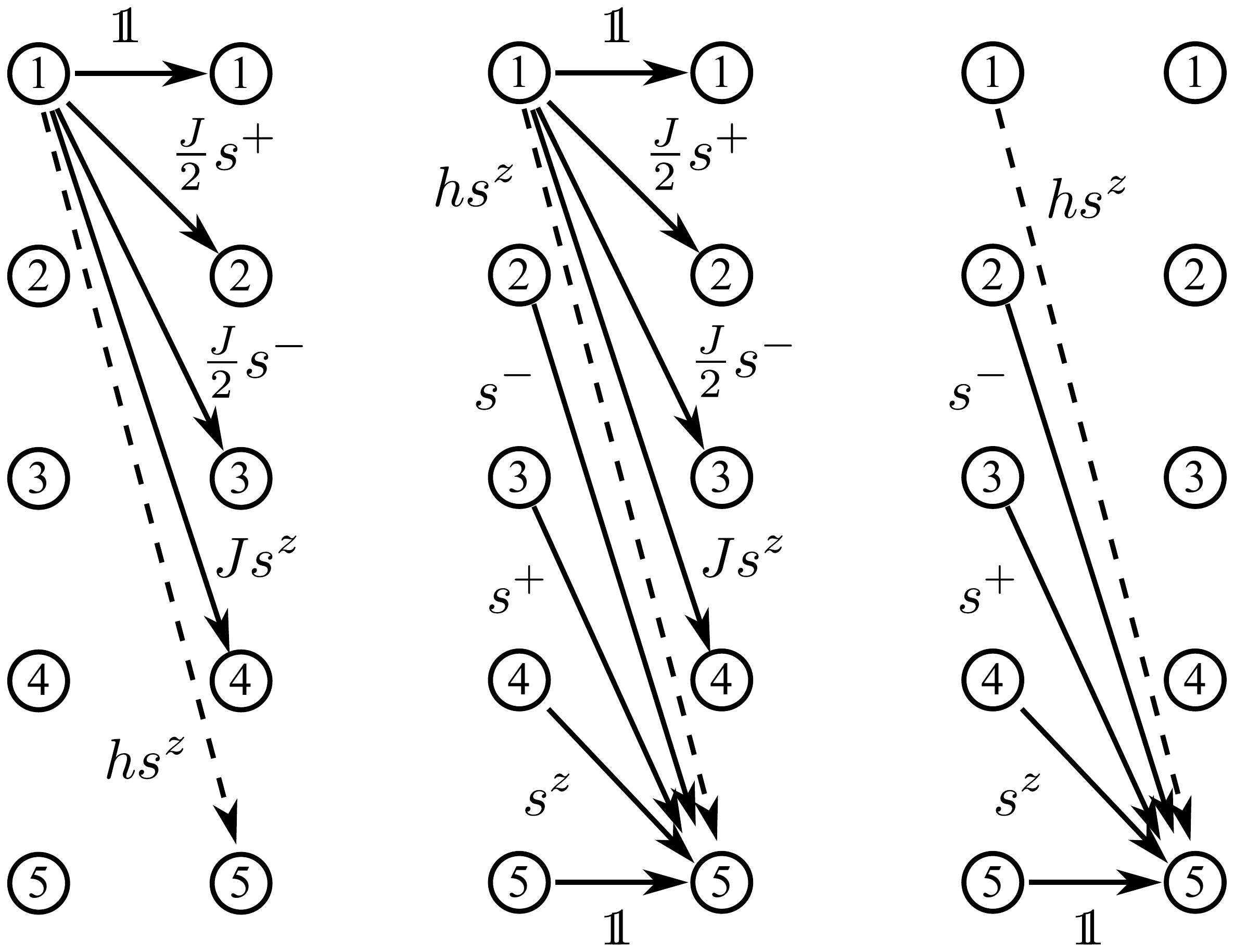}
\caption{The left, centre and right are MP
  diagrammatic forms of Eqns.\ (\ref{eqn:heisW1}), (\ref{eqn:heisWi})
  and (\ref{eqn:heisWL}) respectively. The circles represent the
  virtual indices ($b_{i-1}$, $b_{i}$) of the MPO tensor and arrows
  show the corresponding operator. The dashed arrows highlight the possibility of an additional magnetic field operator $h S^{z}$ not present in (\ref{eqn:heisW1}), (\ref{eqn:heisWi})
  and (\ref{eqn:heisWL}).}
\label{fig:HeisMPdiag}
\end{figure}

\begin{figure}[bt]
\includegraphics[scale=0.3]{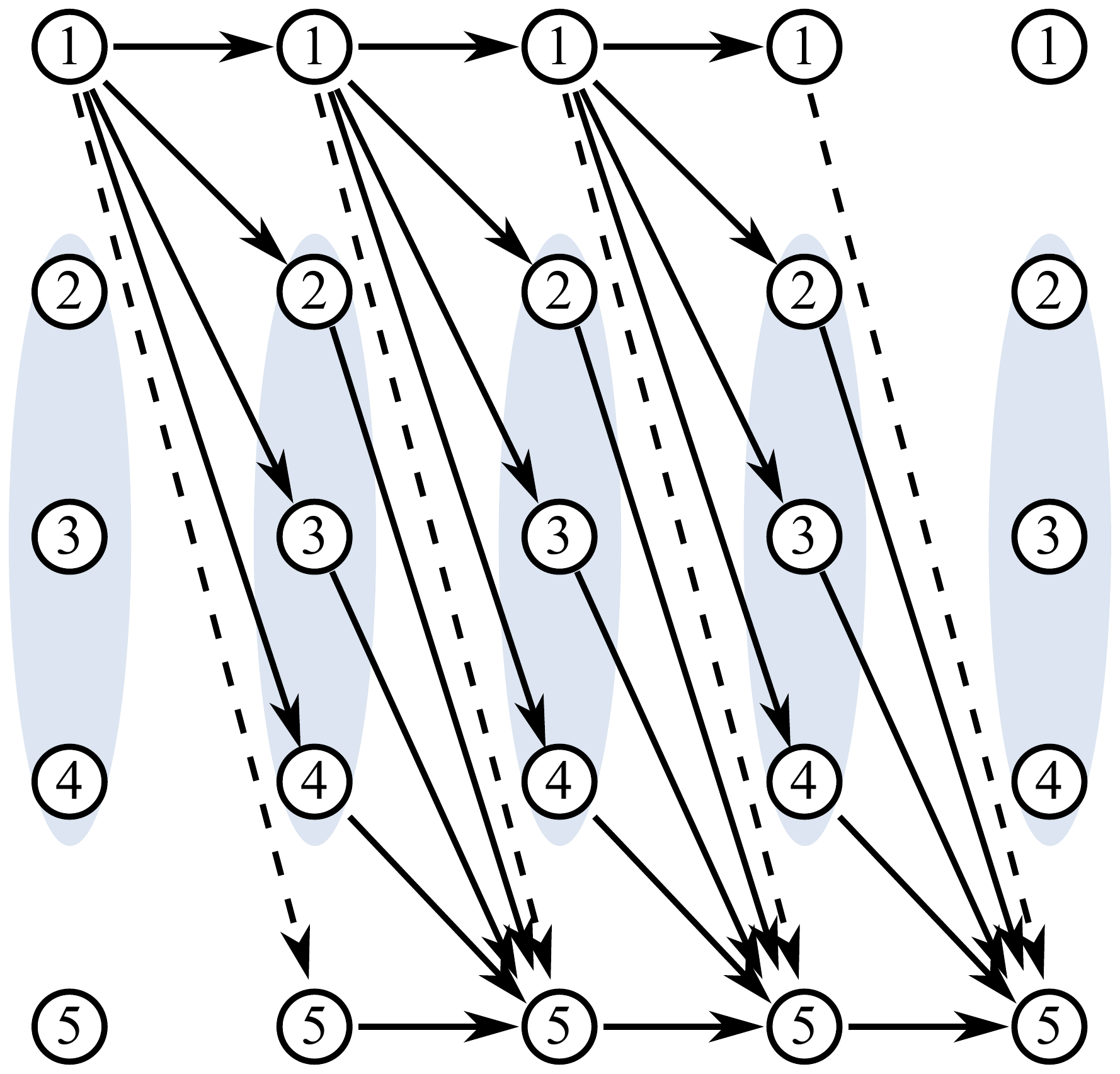}
\caption{(Color online) MP diagram of the contraction of a four-site MPO of the
  Heisenberg Hamiltonian \eqref{eq:hamiltonian}. Symbols and lines as in Fig.\ \ref{fig:HeisMPdiag}. The shaded ellipses linking tensor entries $2$, $3$ and $4$ corresponds to the simplifications employed in Fig.\ \ref{fig:SDRGMPonsite}.}
\label{fig:HeisMPdiag4site}
\end{figure}

Another way of describing the contents of an MPO is a
\emph{matrix product} (MP) diagram.\cite{CroB08} This is a pictorial
representation of the elements in the tensor, whereby the indices are
numbered circles and the corresponding elements are paths that connect
any two indices (Fig.\ \ref{fig:HeisMPdiag}). Matrix multiplication,
or contraction, is then represented by the sum of the unique paths
that connect the indices on the far left and right of the diagram when
multiple matrices are placed end to end (Fig.\
\ref{fig:HeisMPdiag4site}). For MPOs it is understood that the binary operator between terms is a
  tensor product. Tracing out the different paths in Fig.\
\ref{fig:HeisMPdiag4site} results is the standard form \eqref{eq:hamiltonian} with $L=4$. The MP diagrams give a
convenient means of visualising the components of the MPO and are
particularly useful when creating operators with long range components
or periodic boundary conditions.

%

\end{document}